\documentclass[aps,pra,twocolumn,groupedaddress,floatfix,showpacs]{revtex4}
\usepackage{graphicx}
\usepackage{amsmath}
\usepackage{amssymb}
\usepackage{color}
\def\ket#1{|#1\rangle}

\begin{document}
\title{Energy-resolved spin correlation measurements: \\ Decoding transverse spin dynamics in weakly interacting Fermi gases}

\author{J. Huang and J. E. Thomas}

\affiliation{$^{1}$Department of Physics, North Carolina State University, Raleigh, NC 27695, USA}

\date{\today}

\begin{abstract}
We study transverse spin dynamics on a microscopic level by measuring energy-resolved spin correlations in weakly interacting Fermi gases (WIFGs). The trapped cloud behaves as a many-body spin-lattice in energy space with effective long-range interactions, simulating a collective Heisenberg model. We observe the flow of correlations in energy space in this quasi-continuous system, revealing the connection between the evolution of the magnetization and the localization or spread of correlations. This work highlights energy-space correlation as a new observable in quantum phase transition studies of WIFGs, decoding system features that are hidden in macroscopic measurements.
\end{abstract}

\maketitle

Collective spin dynamics plays a central role in spin-lattice models, such as Heisenberg models of quantum magnetism~\cite{QmMagnetism}, Anderson pseudo-spin models of superconductivity~\cite{AndersonPseudoSpin}, and Richardson-Gaudin models of pairing~\cite{RevModPhysRG}. These models have been simulated in discrete systems, including ion traps~\cite{MonroeCorrelProp,ZollerBlattTunableRange,ReyNatPhys2017}, quantum gas microscopes~\cite{QmGasMicroFermi2015}, and cavity-QED experiments~\cite{CavityQEDSmith2016}, which achieve single-site resolution. In contrast, WIFGs provide a powerful many-body platform for realizing spin lattice models in a quasi-continuous system. In the nearly collisionless regime, the energy states of the individual atoms are preserved over experimental time scales, creating a long-lived synthetic lattice~\cite{synthLattice} in energy space that is not achievable in strongly interacting regime. This energy lattice simulates collective Heisenberg Hamiltonians with tunable long-range interactions~\cite{DuSpinSeg2,LewensteinDynLongRange,SaeedPRASpinECorrel,Piechon,MuellerWeaklyInt,LaloeSpinReph,
ThywissenDynamicalPhases,KollerReySpinDep} and adjustable anisotropy \cite{WallEngineerH}.

In this work, we demonstrate measurements of energy-resolved spin correlations, which provide a physically intuitive picture of the transverse spin dynamics in an energy-space spin lattice. This method enables a microscopic look into the signatures of quantum phase transitions and the origins of the macroscopic properties, such as magnetization. In a many-body spin lattice with a collective Heisenberg Hamiltonian, the interplay between the site-dependent energy and site-to-site interactions leads to a transition to a spin-locked state as the interaction strength is increased, producing a large total transverse spin. This transition has been observed in a WIFG of $^{40}$K~\cite{ThywissenDynamicalPhases}, using the total transverse magnetization as the order parameter. More insight into the spin-locking transition is provided by our energy-resolved measurements, which illustrate the emergence of strong correlations between transverse spin components in localized low-energy and high-energy subgroups and the spread of these correlations throughout the energy lattice as interaction strength increases.

The observation of energy-resolved transverse correlation is implemented in a degenerate Fermi gas, consisting of $6.2\times 10^4\,^6$Li atoms. The cloud is confined in an optical trap and cooled to temperature $T=0.21\,T_F$, where Fermi temperature $T_F\approx 0.73\,\mu$K. The ratio between radial and axial trap frequencies is $\omega_r/\omega_x\approx 27$, allowing a quasi-1D approximation for modeling. A superposition of the two lowest hyperfine-Zeeman states, which are denoted by $\ket{\uparrow_z}$ and $\ket{\downarrow_z}$, is prepared by a radiofrequency (RF) pulse at the beginning of each experimental cycle. 

The collision rate is controlled to be negligible during a single cycle by tuning the bias magnetic field $B$ to provide a sufficiently small scattering length $a(B)$. Therefore, in such a weakly interacting regime, the energy and the energy state of each particle are conserved, allowing us to simulate the system as a 1D lattice in energy space. Each lattice site ``$i$" represents the $i^{th}$ harmonic oscillator state along the axial direction of the sample, with an energy  $E_i\!=\!(n_i\!+\!1/2)\,\hbar\omega_x$ and dimensionless collective spin vector $\vec{s}\,(E_i)\equiv\vec{s}_i$. Hence this synthetic lattice can be described by a Heisenberg Hamiltonian \cite{SaeedPRASpinECorrel}:
\begin{equation}
\frac{H(a)}{\hbar}=\sum_{i,j\neq i}\!g_{ij}(a)\,{\vec s}_i\cdot{\vec s}_j+\sum_{i}\Omega'E_i\,s_{zi}.
\label{eq:1.1}
\end{equation}

The first term represents the effective long-range interactions between energy lattice site $i$ and $j$ due to the overlap of probability densities in real space for the energy states $i$ and $j$. $g_{ij}(a)$ is the coupling parameter, scaling linearly with scattering length $a$. The average of $g_{ij}(a)$ for all $ij$ pairs is denoted by $\bar{g}(a)$. 

The second term arises from the magnetic field variation along the axial direction of the cloud, resulting in an effective spin-dependent harmonic potential and corresponding site-dependent detuning rate $\Omega'=-\delta\omega_x/(\hbar\omega_x)$. The statistical standard deviation of $\Omega'\,E_i$, denoted by $\sigma_{\Omega_z}$, determines the spread in the spin-precession rate, $\sigma_{\Omega_z}\approx 1.4$ Hz in our system.

The ratio of these two terms in Eq.~\ref{eq:1.1} determines the behavior of the system during evolution. For this reason, we define the dimensionless interaction strength $\zeta\equiv \bar{g}(a)/(\sqrt{2}\sigma_{\Omega_z})$. Here, larger $\zeta$ represents a stronger mean-field interaction, and for small $\zeta$, the system is dominated by the spread in Zeeman precession.

To predict the dynamics of the system with the Hamiltonian Eq. \ref{eq:1.1}, a quasi-classical spin model is adopted. In the simulation, a mean-field approximation is applied, and the classical collective spin vectors are obtained by neglecting quantum correlations in the Heisenberg equations: $\dot{\vec{s}}_i=[\sum_{j\neq i} g_{ij}(a)\vec{s}_j+\Omega'E_i \hat{e}_z]\times{\vec{s}}_i$~\cite{SaeedPRASpinECorrel,qm_rewind}. The components of collective spin vectors for different energy groups $s_\sigma(E_i)$ are obtained by numerical integration. This equation of motion describes the evolution of spin vectors in the Bloch resonant frame, which rotates at the instantaneous hyperfine resonant frequency for the particles in the lowest energy site: $E_i=0$.

\begin{figure}[hbtp]
\includegraphics[width=0.48\textwidth]{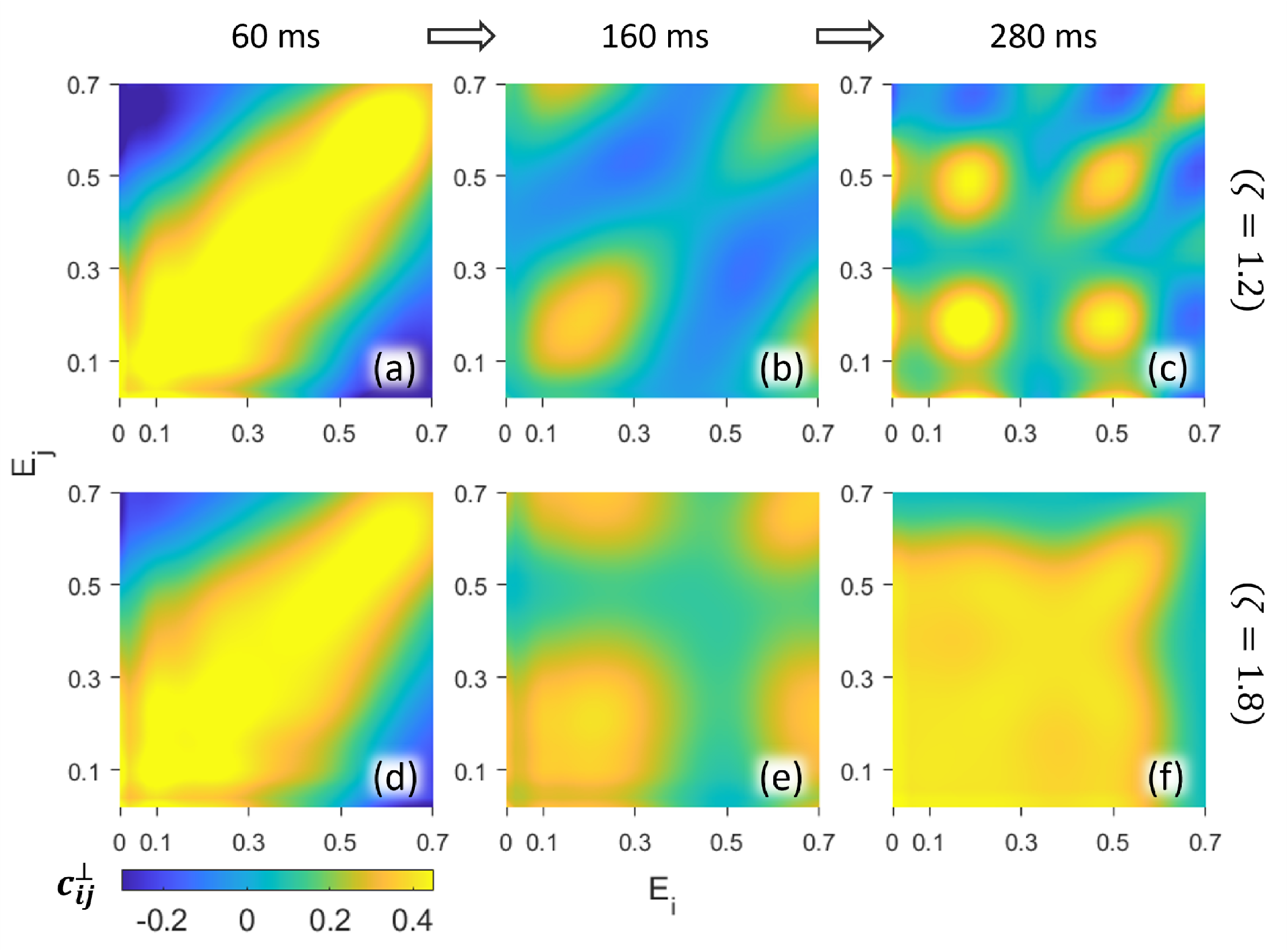}
\caption{Correlation function \textit{\textbf{c}}$^\perp_{ij}$, ensemble-averaged over 30 shots with a selected $\varphi$ distribution, at different evolution times with interaction strength $\zeta=1.2$ (a-c) and $\zeta=1.8$ (d-f). $E_i$ and $E_j$ are in units of effective Fermi energy $E_F$. Only the lowest $70\%$ of energy bins are adopted in data analysis as higher energy groups contain very few particles. The \textit{\textbf{c}}$^\perp_{ij}$ values shown here and Fig. \ref{fig:magnetization_10a}(c-e,h-j) are amplified by dividing by an energy-dependent attenuation coefficient $\Gamma(E_i)$ arising from the finite energy resolution ($\lesssim 0.08\sqrt{E_i}$) to restore the amplitudes to their correct values~\cite{Supplement}.
\label{fig:5a0_spispj}}
\end{figure}

To observe the transverse component of the spin vector, a Ramsey sequence is applied. Starting from an initially $z$-polarized state, the first excitation $(\frac{\pi}{2})_y$ RF pulse produces an $x$-polarized sample. After that, the system is allowed to evolve for a period $\tau$ at the scattering length $a$ of interest. Then, a second $(\frac{\pi}{2})_y$ RF pulse is applied to collectively rotate the spin vectors about the $y$-axis, projecting the $x$-component onto the measurement $z$-axis, ideally. Immediately after the last RF pulse, spin states $\ket{\uparrow_z}$ and $\ket{\downarrow_z}$ are imaged. In reality, as discussed below, $s_z(x)=(n_\uparrow(x)-n_\downarrow(x))/2\equiv s^{meas}(x)$ measures a combination of transverse components of the spin vector in the Bloch resonant frame, $\tilde{s}_{x}$ and $\tilde{s}_y$, just prior to imaging. 

In this quasi-continuous spin system, which contains a large number of atoms with closely spaced energy levels, the spin profiles in real space and in energy space are related by \cite{Supplement}:
\begin{equation}
\label{eq:nEtonX_raw_main}
    s^{meas}(x)=\frac{1}{\pi}\int dE\,|\phi_E(x)|^2\,s^{meas}(E).
\end{equation}
$|\phi_E(x)|^2$ is the probability density which is evaluated using a WKB approximation.
Using Abel inversion, Eq. \ref{eq:nEtonX_raw_main} yields the energy-resolved spin density $\{s^{meas}(E)|E\in[0,E_F]\}$ from measurements in real space $\{s^{meas}(x)|x\in[-\sigma_{Fx},\sigma_{Fx}]\}$~\cite{NewAbelInversion,Supplement}. $E_F$ is the effective Fermi energy and $\sigma_{Fx}$ is the fitted Thomas-Fermi width of the cloud. In the data analysis, we use an energy bin width of $\Delta E=E_F/50$ limited by imaging resolution and the mapping algorithm.

During the experimental cycle, magnetic field fluctuation, at even $10^{-4}$ G level, causes imperfectly controlled RF detuning and subsequent phase $\varphi $ accumulation, changing the relative contribution of the $x$ and $y$ components of spin vectors in the measurement, $s^{meas} = \cos(\varphi )\tilde{s}_x+\sin(\varphi )\tilde{s}_y$. With a broad spread $\varphi \in [0,2\pi]$, a multi-shot average $\langle s^{meas}\rangle$ tends to vanish. As the $\varphi$ distribution for each data set is usually irreproducible, the contribution of the $x$ and $y$ components in $\langle s^{meas}\rangle$ cannot be controlled efficiently and reliably, even with data selection~\cite{Supplement}. 

In the analysis of $s^{meas}$ correlations presented in this work, this problem is circumvented. The correlation between measured operators with energy $E_i$ and $E_j$ has the form~\cite{Supplement}: 
\begin{equation}        
\label{eq:sp_corr}
\begin{split}
     \mathcal{C}^{\perp}_{ij}\equiv\langle s^{meas}_is^{meas}_j\rangle=&\frac{1}{2}\langle \tilde{s}_{xi}\tilde{s}_{xj}+\tilde{s}_{yi}\tilde{s}_{yj}\rangle\\
     +&\frac{1}{2}\langle \cos(2\varphi )\rangle\langle \tilde{s}_{xi}\tilde{s}_{xj}-\tilde{s}_{yi}\tilde{s}_{yj}\rangle\\
     -&\frac{1}{2}\langle \sin(2\varphi )\rangle\langle \tilde{s}_{xi}\tilde{s}_{yj}+\tilde{s}_{yi}\tilde{s}_{xj}\rangle,
\end{split}
\end{equation}
where $\langle\cdots\rangle$ denotes an average over multi-shots, and $\tilde{s}_{\sigma i}$ is the $\sigma$ component of spin vector in the Bloch frame before the last $(\frac{\pi}{2})_y$ pulse. In the data analysis, a data group is selected with a specific phase distribution~\cite{Supplement} to enforce $\langle\cos(2\varphi )\rangle=\langle\sin(2\varphi )\rangle=0$, estimated using the quasi-classical spin model. This method ensures that the correlation obtained by averaging the selected data is $\mathcal{C}^\perp_{ij}= \frac{1}{2}\langle \tilde{s}_{xi}\tilde{s}_{xj}+\tilde{s}_{yi}\tilde{s}_{yj}\rangle$, without making assumptions about the $\varphi $ distribution for the whole data set.

\begin{figure*}[hbtp]
\begin{center}\
\hspace*{-0.15in}
\includegraphics[width=\textwidth]{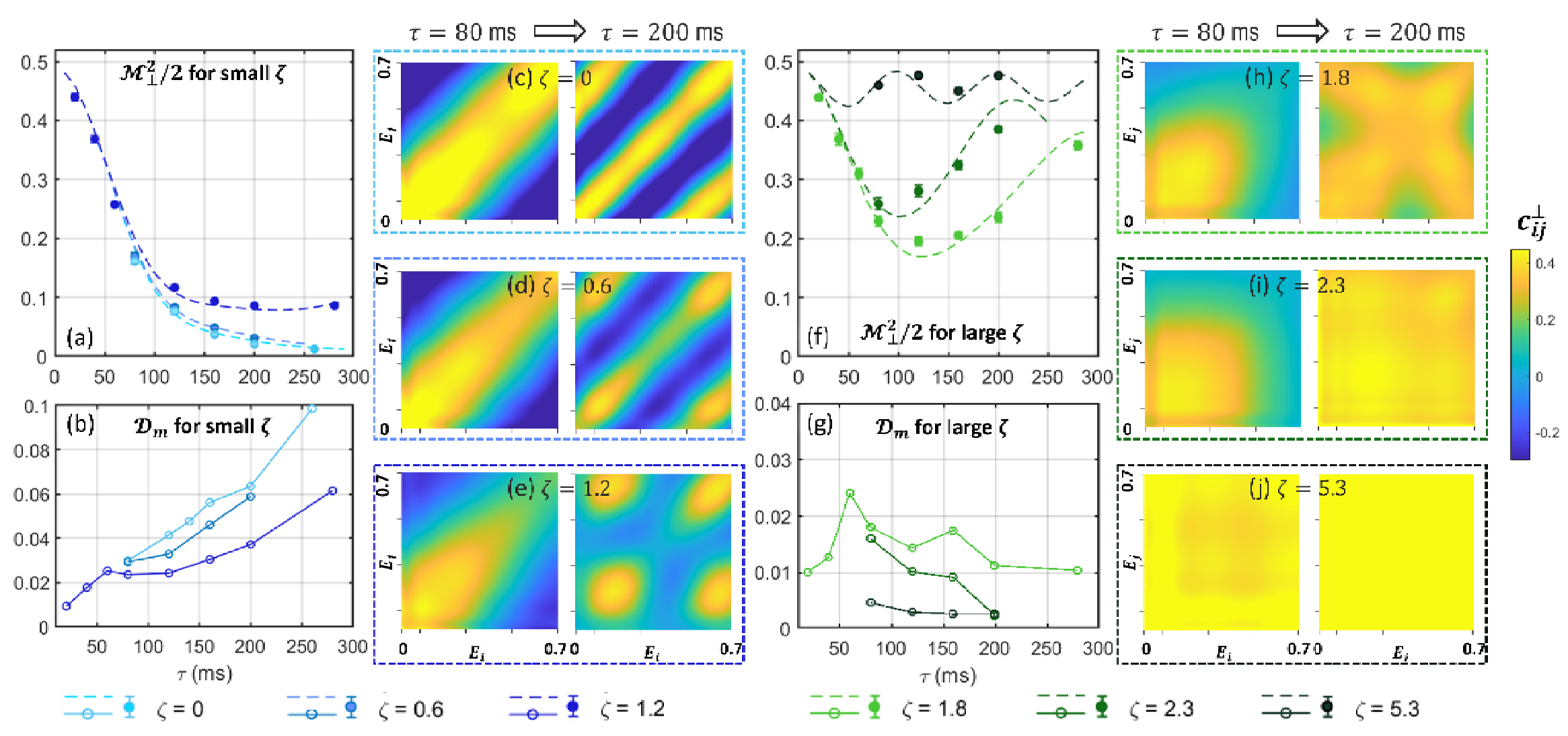}
\end{center}
\caption{Time-dependent transverse magnetization ($\frac{1}{2}\mathcal{M}_\perp^2$) and correlation gradient ($\mathcal{D}_m$) at different interaction strengths, $\zeta$, along with corresponding \textit{\textbf{c}}$_{ij}^\perp$ correlation plots. Solid circles in (a) and (f) are $\frac{1}{2}\mathcal{M}_\perp^2$ obtained by ensemble averaging over multiple shots with the desired $\varphi$ distribution. Darker blue or green corresponds to the cases with stronger interaction. A detailed description of data selection and error bar calculation is in Appendix \ref{sec:data}~\cite{Supplement}. Dashed lines are predictions from the quasi-classical spin model \cite{SaeedPRASpinECorrel,qm_rewind}. Hollow circles in (b) and (g) are correlation gradient $\mathcal{D}_m$ extracted from normalized correlation \textbf{\textit{c}}$_{ij}^\perp$ data at corresponding interaction strength and evolution time. Note that the vertical scale in (g) is expanded to show the details. Correlation plots (c-e,h-j) show \textit{\textbf{c}}$_{ij}^\perp$ at $\tau=80$ ms (left of each pair) and $200$ ms (right of each pair). (c) $\zeta=0\,(a=0\,a_0)$, (d) $\zeta=0.6\,(a=2.62\,a_0)$, (e) $\zeta=1.2\,(a=5.19\,a_0)$, (h) $\zeta=1.8\,(a=8.05\,a_0)$, (i) $\zeta=2.3\,(a=10.54\,a_0)$, (j) $\zeta=5.3\,(a=23.86\,a_0)$. 
\label{fig:magnetization_10a}}
\end{figure*}

In contrast, the longitudinal spin vectors and its correlation, $\langle\tilde{s}_{zi}\rangle$ and $\langle\tilde{s}_{zi}\tilde{s}_{zj}\rangle$, can be measured easily without data selection, as this measurement does not require the last $(\frac{\pi}{2})_y$ RF pulse, and therefore, is insensitive to the RF detuning. We have conducted ensemble averaged $\tilde{s}_z$ measurement and found that  $(\langle\tilde{s}_{zi}\tilde{s}_{zj}\rangle-\langle\tilde{s}_{zi}\rangle\langle\tilde{s}_{zj}\rangle)/(N_iN_j/4)$ has a value of $\sim 5\times10^{-3}$, which is comparable to spin projection noise, indicating the system is not quantum correlated. In addition, as our previous single-shot measurements showed, this large spin system can be well explained by the quasi-classical model~\cite{qm_rewind}. Therefore, this system is expected to evolve classically, where the classical correlation $\mathcal{C}_{ij}^\perp$ is of interest. By construction, $\mathcal{C}_{ij}^\perp$ also detects quantum correlations when they are present.

To study the correlation between one pair of particles with energies $E_i$ and $E_j$, $\mathcal{C}_{ij}^\perp$ is normalized by atom numbers in the $i^{th}$ and $j^{th}$ energy partitions, $N_i$ and $N_j$. The normalized transverse correlation is defined as: \textbf{\textit{c}}$_{ij}^\perp\equiv\mathcal{C}_{ij}^\perp/(\frac{N_jN_j}{4})$. Then by construction from Eq.~\ref{eq:sp_corr}, \textbf{\textit{c}}$_{ij}^\perp\in [-\frac{1}{2},\, \frac{1}{2}]$. In this work, it is observed that the normalized transverse correlation evolves in qualitatively distinct ways as the interaction strength, $\zeta$, increases. 

Fig. \ref{fig:5a0_spispj} illustrates the different behaviors of \textbf{\textit{c}}$_{ij}^\perp$ at $\zeta=1.2$ (top row (a-c)) and $\zeta=1.8$ (bottom row (d-f)). At early time, for both interaction strengths, the spins are x-polarized and the transverse spin components are mostly self correlated, and their \textbf{\textit{c}}$_{ij}^\perp$ have very similar distributions in energy space as shown in Fig. \ref{fig:5a0_spispj}(a,d). As time evolves, in the system with smaller interaction strength (Fig. \ref{fig:5a0_spispj}(b,c)), the single particle pair correlation tends to be localized between multiple specific energy subgroups. In contrast, for the case with stronger interaction (Fig. \ref{fig:5a0_spispj}(e,f)), the correlation tends to become more uniform across all pairs of energy groups at a later time. This distinct behavior of microscopic correlations reveals the source of the phase transition that the system undergoes. 

In addition to visualizing the distribution of highly correlated regions in the energy lattice using surface plots (Fig.~\ref{fig:5a0_spispj}), the energy resolved transverse correlation measurement directly yields the macroscopic transverse magnetization, which undergoes a phase transition as interaction strengths increase. The system magnetization is related to the ensemble-averaged correlation functions. The square of total transverse magnetization $\mathcal{M}_\perp^2=S_x^2+S_y^2$ is the double summation of the transverse correlation in energy space: $\frac{1}{2}\mathcal{M}^2_\perp=\sum_{i,j}\mathcal{C}_{ij}^\perp$. In this way, $\frac{1}{2}\mathcal{M}^2_\perp $ data presented in this work are calculated using our measured $\mathcal{C}_{ij}^\perp$. However, the macroscopic magnetization does not fully represent the structure of the correlations across the energy-space landscape.

\begin{figure*}[hbtp]
\hspace*{-0.25in}
\includegraphics[width=\textwidth]{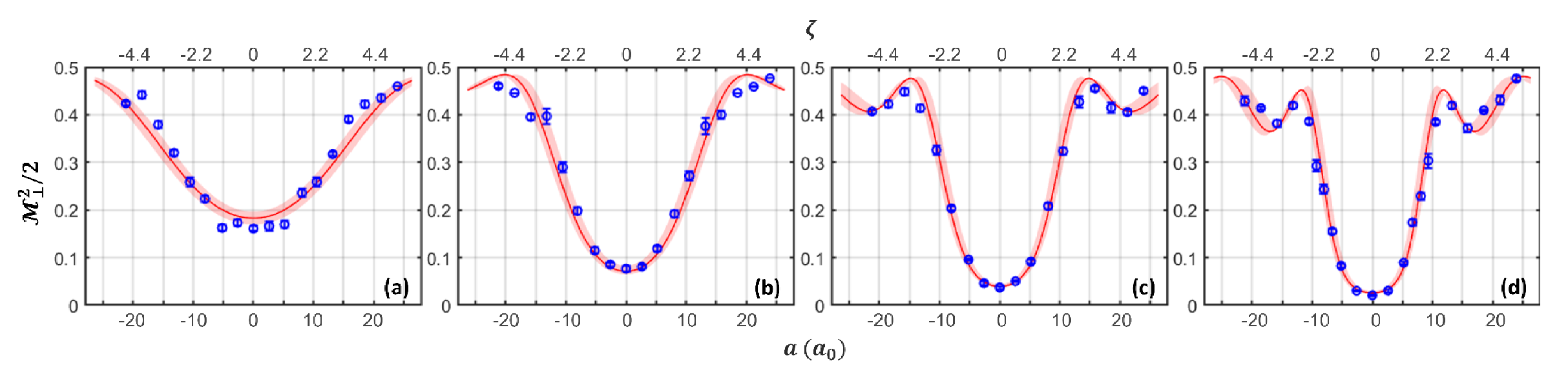}

\caption{Observing the emergence of spin locking by measuring $\frac{1}{2}\mathcal{M}_{\perp}^2$ for various interaction strength $\zeta$ (top axis) and corresponding scattering lengths $a$ (bottom axis) at (a) 80 ms, (b) 120 ms, (c) 160 ms, and (d) 200 ms. Blue circles are averaged data over multiple shots with the same averaging and error bar calculation for Fig. \ref{fig:magnetization_10a}. Bright red curves are predictions with the quasi-classical spin evolution model and the pink bands correspond to a $2\%$ standard deviation in cloud size $\sigma_{Fx}$. 
\label{fig:spinLock_4tau}}
\end{figure*}

The $\mathcal{C}^\perp_{ij}$ measurement opens new ways to observe the microscopic spin dynamics of the system in energy space. One method to describe the microscopic information is to quantify the extent of the correlations, by determining the magnitude of the correlation-gradient near the point of maximum correlation, $\mathcal{D}_m$.
To calculate $\mathcal{D}_m$ from the correlation matrix \textbf{\textit{c}}$_{ij}^\perp$, first, we find the energy partition $E_i=E_m$ where the center of the highest correlated region is located. Then we calculate the absolute values of the gradient for the transverse spin correlation between this energy partition and all other partitions, \textbf{\textit{c}}$_{mj}^\perp$. Finally, $\mathcal{D}_m$ is defined as the average magnitude of the gradient of \textbf{\textit{c}}$_{mj}^\perp$ for all energy partitions $j\in [1, j_{max}]$ for the fixed $m$:
\begin{equation}
    \mathcal{D}_m \equiv\frac{1}{j_{max}}\sum_{j=1}^{j_{max}}|\nabla \textit{\textbf{c}}_{mj}^\perp|,
\end{equation}
where $j_{max}$ is the number of total energy groups adopted in data analysis \cite{Supplement}. Therefore, $\mathcal{D}_m$ measures the maximum magnitude of the gradient for normalized correlations between transverse spin vectors in one energy partition and in all other partitions. A large $\mathcal{D}_m$ value indicates that high correlations cluster around specific energy group pairs $E_m$ and some of $E_j$. A small $\mathcal{D}_m$ means that the high correlation region is spread evenly across most lattice site pairs in energy space. 


The time-evolution of $\mathcal{M}^2_\perp$ and $\mathcal{D}_m$ at different interaction strengths is shown along with corresponding microscopic transverse correlation plots in Fig. \ref{fig:magnetization_10a}. In this figure, panels in the left half (a-e) for small interaction strength $\zeta$ and panels in the right half (f-j) for large interaction strength demonstrate two distinct behaviors. A system with small interaction strength ($\zeta = 0,\,0.6,\,1.2$) tends to demagnetize as time evolves: $\frac{1}{2}\mathcal{M}^2_\perp(t\rightarrow\infty)$ asymptotes to a small value (Fig. \ref{fig:magnetization_10a}(a)). The normalized transverse correlation in such a system acts similarly to the example in the top row of Fig. \ref{fig:5a0_spispj}: the largest correlations $|$\textbf{\textit{c}}$_{ij}^\perp|$ (either positive or negative) arise between certain localized energy groups, either forming thin stripes or forming islands, as shown in Fig. \ref{fig:magnetization_10a} (c-e). The corresponding maximum correlation gradient at these interaction strengths increases over time (Fig. \ref{fig:magnetization_10a}(b)), in agreement with the features of surface plots (c-e). 
In contrast, for stronger interactions ($\zeta\geq 1.8$), $\frac{1}{2}\mathcal{M}^2_\perp(t\rightarrow\infty)$ oscillates relative to a larger static level as $\zeta$ increases (Fig. \ref{fig:magnetization_10a}(f)). In such cases, (h-j) suggests that the high correlation domain tends to extend over all pairs of energy lattice sites, as opposed to the trend in (c-e). The measurement of $\mathcal{D}_m$ illustrates this trend in a quantitative way: \textbf{\textit{c}}$_{ij}^\perp$ has a persistent low correlation gradient (Fig. \ref{fig:magnetization_10a}(g)), corresponding to an extended correlation region.

Furthermore, even when $\mathcal{M}_\perp^2$ has the same value at two different times, by comparing the corresponding correlation plots, it is observed that the strongly correlated region in energy space can have completely different distributions. For example, for $\zeta=1.8$ (the lightest green data) in Fig.~\ref{fig:magnetization_10a}(f), $\mathcal{M}_\perp^2(80\,\text{ms})=\mathcal{M}_\perp^2(200\,\text{ms})$ , but the corresponding \textbf{\textit{c}}$_{ij}^\perp$ (Fig.~\ref{fig:magnetization_10a}(h)) shows different features for these two times: at $80$ ms, the transverse spin vectors are strongly correlated mainly between low energy groups, and in contrast, at $200$ ms, the high transverse correlation domain has extended to energy partition pairs that are further apart. Similarly, for $\zeta=1.2$ (the darkest blue data) in Fig.~\ref{fig:magnetization_10a}(a), $\mathcal{M}_\perp^2(200\,\text{ms})=\mathcal{M}_\perp^2(280\,\text{ms})$, but Fig.~\ref{fig:magnetization_10a}(e) and Fig.~\ref{fig:5a0_spispj}(c) show different distributions of \textbf{\textit{c}}$_{ij}^\perp$. These different structures observed in correlation plots are very well represented by corresponding high and low values of $\mathcal{D}_m$ for these cases. Therefore, the observations of energy-resolved transverse correlation provide new probes to characterize the spin dynamics more deeply than simply measuring macroscopic quantities.

From the measured energy-space correlation function \textbf{\textit{c}}$_{ij}^\perp$, we conclude that a system with a more localized transverse correlation between multiple specific energy group pairs tends to be demagnetized as time evolves (Fig.~\ref{fig:magnetization_10a}(a)). In contrast, a system with the transverse correlation spread over most energy lattice site pairs at a long evolution time maintains the high initial magnetization (Fig.~\ref{fig:magnetization_10a}(b)). These transitions in the magnetization with increasing interaction strength are shown in Fig.~\ref{fig:spinLock_4tau} for four evolution times. Blue circles are $\mathcal{M}^2_\perp$ obtained directly from the double sum of the correlation function as described above. Predictions of $\mathcal{M}_\perp^2$ (red curves) are obtained using the quasi-classical model. We find that, as the interaction strength increases, the transverse magnetization surges, simulating the transition from a paramagnetic phase to a ferromagnetic phase. Fig. \ref{fig:magnetization_10a}(c-e,h-j)  shows how the corresponding spin correlations change from localized to global across this transition. 

In summary, we have developed energy-space spin correlation measurement as a method for characterizing the spin dynamics of quasi-continuous systems, which simulate a synthetic lattice of spins pinned in energy space. This method enables a full microscopic view of how correlations develop between the extensive subsets of spins in energy space, associating the evolution of the macroscopic properties with the local correlation behavior. Utilizing this idea, we connect the spread and localization of correlations to the system magnetization and demagnetization by observing the correlation distribution as a function of time and interaction strength. 

The new observables developed in this work are broadly applicable in weakly interacting quantum gases. In these systems, long-range interactions between lattice sites in energy space can be engineered to simulate a wide variety of model Hamiltonians. For example, tunable spatial asymmetry can be introduced into the coupling constant by creating spin-dependent energy states \cite{WallEngineerH}. Further, the scattering length can be controlled with high spatial and temporal precision with optical control technique \cite{JaganathanOptControl}. Therefore, these energy-resolved probes can be exploited in broad studies of macroscopic out-of-equilibrium dynamics and critical dynamics across quantum phase transitions in quantum simulators.


We thank Ilya Arakelyan for helpful discussions. Primary support for this research is provided by the Air Force Office of Scientific Research (FA9550-22-1-0329). Additional support is provided by the National Science Foundation (PHY-2006234 and PHY-2307107).

$^*$Corresponding authors: jhuang39@ncsu.edu\\ \hspace*{1.65 in} jethoma7@ncsu.edu


%

\widetext
\setcounter{figure}{0}
\setcounter{equation}{0}
\renewcommand{\thefigure}{S\arabic{figure}}
\renewcommand{\theequation}{S\arabic{equation}}

\appendix
\section{Supplemental Material}

This supplemental material presents details of experimental procedures, data analysis, and modeling for measurement of transverse spin components in a weakly interacting Fermi gas. We discuss the mathematical formalism of Abel inversion and apply it to obtain the energy-resolved spin density from measurement in real space. A data selection method specific to measurement of transverse spin correlation is described and illustrated by deriving the final state of a non-interacting gas after a Ramsey sequence. Finally, an energy-dependent attenuation coefficient for the ensemble-averaged correlation, arising from the finite energy resolution, is introduced and tested on data.

\maketitle

\subsection{Experimental procedure}
The experiment presented in this work is implemented with $^6$Li atoms. A 50-50 incoherent mixture of two lowest hyperfine states $\ket{1}$ and $\ket{2}$ are evaporatively cooled in a CO$_2$ laser optical trap to degeneracy. Then state $\ket{1}$ is eliminated with a $17\,\mu$s imaging pulse at a magnetic field of $\sim1200$G. With one spin state left in the sample, the magnetic field is swept close to zero-crossing (527.15 G) so that both the magnitude and sign of scattering length can be tuned during experiments. After the magnetic field is stabilized at the experimental value, an excitation RF $(\frac{\pi}{2})_y$ (0.5 ms), which is on resonance with $\ket{2}$ to $\ket{1}$ transition, is applied to create an $x$-polarized sample. Following the RF excitation, the system is allowed to evolve with s-wave scattering for a time period $\tau$. Then another $(\frac{\pi}{2})_y$ RF pulse is applied to observe the transverse components of spin vectors. Immediately after the Ramsey sequence is completed, two imaging pulses separated by $10\,\mu$s are shined on the sample to obtain the absorption images of both spins. With these images, integration across the radial direction yields the axial spatial density profiles $n_1(x)$ and $n_2(x)$. With the technique of Abel inversion (introduced in \S\ref{sec:Abel}), the energy space profiles $n_1(E)$ and $n_2(E)$ of the two states are obtained from the measured spatial profiles.

\subsection{Abel Inversion}
\label{sec:Abel}
In this section, we introduce a numerical solution of Abel-type integral equations using an improved method that was first presented in 1991~\cite{NewAbelInversion}. We apply this method to extract energy space spin densities from spatial profiles measured in the experiments presented in this work. This Abel inversion solution is obtained by expanding the energy integral in a series of cosine functions whose amplitudes are calculated by least-squares-fitting from the measured spatial data. The number of expansion terms is determined by the complexity of the data. We use 8-12 expansion terms for the data presented in this work.

\subsubsection{Formalism}
Abel inversion is an optimum way to evaluate the energy distribution of a given spatial profile along the longitudinal axis, although it was first designed for the extraction of a distribution in the radial direction from a measurement of distribution along one axis $y$. The Abel transform has the form
\begin{equation}
\label{eq:ApdxAbelTrans}
    h(y)=2\int^R_y f(r)\frac{r\,dr}{\sqrt{r^2-y^2}},
\end{equation}
where $h(y)$ is a measured physical quantity and $f(r)$ is an unknown function. To evaluate $f(r)$, it is expanded into $n_{max}$ cosine terms~\cite{NewAbelInversion}:
\begin{equation}
\label{eq:ApdxAbelPaperf_r}
    f(r) = \sum_{n=0}^{n_{max}}A_n\,f_n(r),
\end{equation}
\begin{equation}
\label{eq:Abel_cos_expansion}
    f_n(r)= \left\{\begin{array}{lcl}
    0 & \text{ for }n=0;\\
    1-(-1)^n \cos\left(n\pi\frac{r}{R}\right)& \text{ otherwise}.
    \end{array}\right.
\end{equation}

Using Eq. \ref{eq:ApdxAbelTrans} and the Abel transform of Eq. \ref{eq:ApdxAbelPaperf_r}, we define $H(y)$ as a summation of $n_{max}$ terms:

\begin{equation}
\label{eq:AbelExpand1}
\begin{split}
    H(y)=& \,2\sum_{n=0}^{n_{max}}A_n\int_y^Rf_n(r)\frac{r\,dr}{\sqrt{r^2-y^2}}\\
    =&\,2\sum_{n=0}^{n_{max}}A_nh_n(y)
\end{split}
\end{equation}
where $y$ is the independent variable.

Assume that in each measurement, there are $k_{max}$ values of $y_k$ in total, forming set $Y$. Therefore, the corresponding measured value $d_k=h(y_k)$ has $k_{max}$ values, too, forming set $D$: 

\begin{equation}
\begin{split}
    h|_Y:Y\rightarrow D; \qquad & 
    d_k\in D:\{k\in \mathbb{N}_1:\,k\leq k_{max}\}\\
    &y_k\in Y:\{k\in \mathbb{N}_1:\,k\leq k_{max}\}.
\end{split}
\end{equation}
In this work, the independent variable $x$ for the spatial profile $s^{meas}(x)$ is scaled by the size of the cloud $\sigma_{Fx}$ for each shot such that $x_k/\sigma_{Fx}\in[-1,1]$. Then the raw measured axial spatial profile $s^{meas}(x)$ is folded over the center of cloud where $x_k=0$, and binned into 50 points. After this process, $x_k/\sigma_{Fx}\in[0,1]$. It's convenient to use: $y_k\rightarrow x_k/\sigma_{Fx}$, thus $y_k\in[0,1]$ and $k_{max}=50$.

With this measurement, calculate the squared difference between expansion $H(y)$ and measured $h(y)$:
\begin{align*}
      \chi^2&=\sum_{k}[h(y_k)-H(y_k)]^2\\
      &=\sum_{k}\left[d_k-\sum_{n=0}^{n_{max}}A_nh_{n,k}\right]^2
\end{align*}
and minimize it by $\frac{\partial \chi^2}{\partial A_n}=0$
\begin{align}
\label{eq:AbelFindA}
    &2\sum_{k}\left[d_k-\sum_{n=0}^{n_{max}}A_nh_{n,k}\right]h_{m,k}=0\\
    &\underbrace{\sum_k d_kh_{m,k}}_{\equiv V_m}=\underbrace{\sum_n\sum_k h_{n,k}h_{m,k}}_{\equiv M_{mn}}A_n\\
    &\Rightarrow\qquad A_n=(M^{-1})_{nm}V_m.
\end{align}
With the values of coefficients $A_n$ optimized, the expansion $H(y)$ will have a numerical form.

\subsubsection{Real space and energy space}
\label{sec:AspatialProfile}
To apply the Abel inversion to obtain energy profiles from spatial ones, the correspondence between them needs to be understood. In this section, we illustrate how to achieve the form shown in Eq. \ref{eq:ApdxAbelTrans} with correct dimensions.

Atom density in real space and energy space is connected by the density probability function:
\begin{equation}
\label{eq:nEtonX_raw}
    n_\sigma(x)=\frac{1}{\pi}\int dE\,|\phi_E(x)|^2\,n_\sigma(E).
\end{equation}
A WKB approximation is applied to evaluate $|\phi_E(x)|^2$:
\begin{align*}
    |\phi_E(x)|^2 \approx \frac{\Theta\left[\frac{2E}{m\omega_x^2}-x^2\right]}{\pi\sqrt{\frac{2E}{m\omega_x^2}-x^2}},\\
    \Rightarrow n_\sigma(x)=\frac{1}{\pi}\int_{\frac{m\omega_x^2 x^2}{2}}^{E_F}\frac{dE\,n_\sigma(E)}{\sqrt{\frac{2E}{m\omega_x^2}-x^2}},
\end{align*}
where $E_F=\frac{1}{2}m\omega_x^2\sigma_{Fx}^2$ is an effective Fermi energy, and the cutoff $\sigma_{Fx}$ is obtained by fitting a zero temperature Thomas Fermi profile to the spatial density of the sample.

For convenience in the calculations, the variables are converted to dimensionless forms:
\begin{align*}
    \tilde{E}\equiv\frac{E}{E_F}, \qquad d\tilde{E}=\frac{dE}{E_F}.
\end{align*}
Then the upper and lower limits of the integral become:
\begin{equation}
\left\{
\begin{array}{cl}
     \text{upper limit }&\rightarrow 1 \\
     \text{lower limit }&\rightarrow \frac{m\omega_x^2x^2}{2}/E_F = \frac{x^2}{\sigma_{Fx}^2}.
\end{array}
\right.    
\end{equation}
\begin{equation}
    \Rightarrow n_\sigma(x)=\frac{1}{\pi}\int^1_{x^2/\sigma_{Fx}^2}\frac{d\tilde{E}\,n_\sigma(\tilde{E})}{\sigma_{Fx}\sqrt{\tilde{E}-\frac{x^2}{\sigma_{Fx}^2}}}.
\end{equation}
Now the spatial variable is changed to its dimensionless version:
\begin{align*}
    \tilde{X}\equiv \frac{x}{\sigma_{Fx}}
\end{align*}

\begin{equation}
\label{eq:ApdxXtoEprofileDmsls}
    \Rightarrow n_\sigma(\tilde{X})=\frac{1}{\pi\sigma_{Fx}}\int^1_{\tilde{X}^2} \frac{d\tilde{E}\,n_\sigma(\tilde{E})}{\sqrt{\tilde{E}-\tilde{X}^2}}.
\end{equation}
Eq. \ref{eq:ApdxXtoEprofileDmsls} has the same form as Eq. \ref{eq:ApdxAbelTrans} by writting
$$\frac{r}{R}\rightarrow\sqrt{\tilde{E}},$$
and the expansion terms in Eq.~\ref{eq:Abel_cos_expansion} has the form:
\begin{equation}
\label{eq:Abel_cos_expansion2}
    f_n(\sqrt{\tilde{E}})= \left\{\begin{array}{lcl}
    0 & \text{, for }n=0;\\
    1-(-1)^n \cos\left(n\pi\sqrt{\tilde{E}})\right)& \text{, otherwise}.
    \end{array}\right.
\end{equation}
$h_n(y)$ in Eq.~\ref{eq:AbelExpand1} becomes:
\begin{equation}
    \label{eq:AbelExpand2}
    h_n(\tilde{X})=\frac{1}{\sigma_{Fx}}\int^1_{\tilde{X}^2} \frac{d\tilde{E}}{\sqrt{\tilde{E}-\tilde{X}^2}}f_n(\sqrt{\tilde{E}}).
\end{equation}

\begin{figure}[hbtp]
\begin{center}
\includegraphics[width=0.6\textwidth]{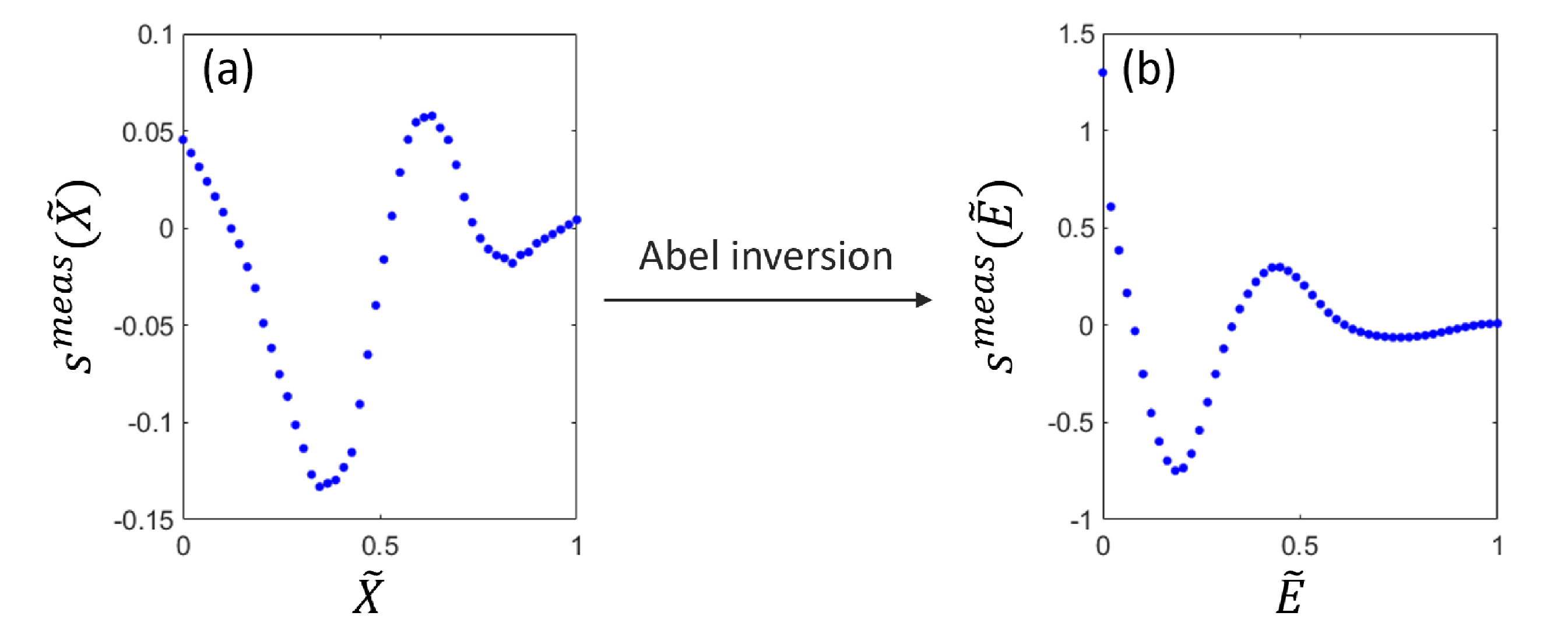}
\end{center}
\caption{Mapping real space spin density measurement (a) to energy space (b) using Abel inversion. 
\label{fig:AbelInv_EG}}
\end{figure}

Performing least-squares-fitting, as introduced in Eq.~\ref{eq:AbelFindA}, with data $d_k=n_\sigma(\tilde{X}_k)$ and Eq.~\ref{eq:AbelExpand2}, will yield a numerical form of $n_\sigma(\tilde{E})$. In this way, the spin density in energy space is obtained. Fig. \ref{fig:AbelInv_EG} shows an example of single-shot measurement of the spin vector in real space (a) and the energy space spin density (b) obtained by the Abel inversion method introduced in this section.

\subsubsection{Energy resolution}

The imaging system in our lab has a spatial resolution $\delta x\approx5\,\mu$m. This results in a finite resolution in energy space, $\delta E_s$. Since the energy is related to the spatial position, 
$$E=\frac{1}{2}m\omega_x^2x^2,$$
the resolution in energy space can be estimated by:
\begin{equation}
\label{eq:ds}
\delta E_s\approx \delta(\frac{1}{2}m\omega_x^2x^2)=\frac{2x\delta x}{\sigma_x^2} E_F \approx 0.03\, E_F\sqrt{\tilde{E}}\, ,
\end{equation}
since $x^2/\sigma_{Fx}$ scales as $\sqrt{E_i/E_F}$.

Furthermore, $n_\sigma(\tilde{E})$ has a finite resolution in energy space, $\delta E_A$, which is related to the maximum number of terms, $n_{max}$, adopted in the Abel inversion. From Eq.~\ref{eq:Abel_cos_expansion2}, we can write the $n^{th}$ ($n>0$) expansion term as 
$$f_n(\sqrt{\tilde{E}})=1-(-1)^n \cos{\phi'_{n}(\sqrt{\tilde{E}})}\, ,$$ where $\phi'_n(\sqrt{\tilde{E}})\equiv n\pi\sqrt{\tilde{E}}$. Then the resolution $\delta E_{A}$ can be estimated by setting
\begin{equation}
\label{eq:da}
\begin{split}
    1&\simeq \delta E_A\frac{\partial \phi'_n(\sqrt{\tilde{E}})}{\partial E}\Bigg\rvert_{n=n_{max}}\\
    1&=\delta E_A\frac{\partial (n_{max}\pi\sqrt{E_i/E_F})}{\partial{E_i}}\\
    \Rightarrow \delta E_A&=\frac{2}{n_{max}\pi}\sqrt{\frac{E_i}{E_F}}E_F.
\end{split}
\end{equation}
In the data analysis, $n_{max}=12$ is applied to most of the data, yielding $\delta E_{A}\approx 0.06\,E_F\sqrt{\tilde{E}}$.

\subsection{Collective Spin Vector Evolution Model}
To understand the effect of the RF detuning on $s^{meas}(x)$ measurement, we derive the final state of the system after the second $(\frac{\pi}{2})_y$ pulse. Note that, although in the main text, the measured quantity is written as $s^{meas}(x)$, it is really the $z$-component of the spin vector at the moment the system is imaged. However, after the last $(\frac{\pi}{2})_y$ RF pulse, there is no evolution time before imaging. Therefore, the measured quantity contains the transverse components of the spin vector just prior to imaging. For convenience, $s^{meas}$ in the main text is written as $s_z$ in this section so that the rotation can be illustrated more straightforwardly in the derivation. 

Prior to the pulse sequence, the optically trapped atoms are initially prepared in a $z$-polarized state,
\begin{equation}
\ket{\psi_{0z}}=\Pi_i\,\ket{\!\!\uparrow_{zi}}.
\end{equation}
Therefore, the final state after the pulse sequence is
\begin{equation}
    \label{eq:finalstate_raw}
    |\psi_f\rangle = e^{-i\frac{\pi}{2}S_y} e^{-i\frac{H}{\hbar}\tau}\underbrace{e^{-i\frac{\pi}{2}S_y}|\psi_{0z}\rangle}_{|\psi_{0x}\rangle}.
\end{equation}
If include the RF detuning into the Hamiltonian, Eq.~1 in the main text has a form:
\begin{equation}
    \frac{H}{\hbar}=\sum_{i,j\neq i}g_{ij}\vec{s}_i\cdot\vec{s}_j+\sum_i(\Omega'E_i+\Delta(t))s_{zi},
\end{equation}
where $\Delta(t)$ represents the RF detuning rate with unknown time dependence. Keeping the detuning part separate, we define the Hamiltonian with two parts:
\begin{equation}
    \frac{H}{\hbar}=\frac{H^0}{\hbar}+\Delta(t) S_z,
\end{equation}
then $H/\hbar\tau$ becomes:
\begin{equation}
\label{eq:ch7H2parts}
    \frac{H}{\hbar}\tau=\frac{H^0}{\hbar}\tau+\int_\tau \Delta(t)\,dt S_z=\frac{H^0}{\hbar}\tau+\varphi S_z ,
\end{equation}
where $\varphi $ is the accumulated phase shift due to RF detuning for an evolution period $\tau$.

Then the measurement of $s_z$ yields:
\begin{equation}
    \langle\psi_f|s_{zi}|\psi_f\rangle=\langle\psi_{0x}|e^{i\frac{H}{\hbar}\tau}\underbrace{e^{i\frac{\pi}{2}S_y}s_{zi}e^{-i\frac{\pi}{2}S_y}}_{s_{zi}(\frac{\pi}{2})_y}e^{-i\frac{H}{\hbar}\tau}|\psi_{0x}\rangle.
\end{equation}
With differential equation $s''_{zi}(\theta_y)=-s_{zi}(\theta_y)$ and initial conditions $s_{zi}(0_y)=s_{zi}$ and $s_{zi}'(0_y)=-s_{xi}$, we obtain the analytic form:
\begin{equation}
\label{eq:ch5RotY}
   s_{zi}(\theta_y) =\cos(\theta_y)s_{zi}-\sin(\theta_y)s_{xi}.
\end{equation}
Then
\begin{align*}
        \langle\psi_f|s_{zi}|\psi_f\rangle&=\langle\psi_{0x}|e^{i\frac{H}{\hbar}\tau}(-s_{xi})e^{-i\frac{H}{\hbar}\tau}|\psi_{0x}\rangle\\
        &=\langle\psi_{0x}|e^{i\frac{H^0}{\hbar}\tau}e^{i\varphi  S_z}(-s_{xi})e^{-i\frac{H^0}{\hbar}\tau}e^{-i\varphi  S_z}|\psi_{0x}\rangle.
\end{align*}

Since the commutation rule $[H^0, S_z]=0$, rearrange the terms to get:
\begin{equation}
    \langle\psi_f|s_{zi}|\psi_f\rangle=\langle\psi_{0x}|e^{i\frac{H^0}{\hbar}\tau}e^{i\varphi  S_z}(-s_{xi})e^{-i\varphi  S_z}e^{-i\frac{H^0}{\hbar}\tau}|\psi_{0x}\rangle.
\end{equation}
Similar to Eq.~\ref{eq:ch5RotY}, we can derive 
\begin{equation}
\label{eq:Sxofvarphi}
   s_{xi}(\theta_z) \equiv e^{i\theta S_z}s_{xi}e^{-i\theta S_z}=\cos(\theta_z)s_{xi}-\sin(\theta_z)s_{yi}.
\end{equation}
Then the measurement of $s_{zi}$ becomes:
\begin{equation}    \langle\psi_f|s_{zi}|\psi_f\rangle=\langle\psi_{0x}|e^{i\frac{H^0}{\hbar}\tau}(s_{xi}\cos\varphi -s_{yi}\sin\varphi )e^{-i\frac{H^0}{\hbar}\tau}|\psi_{0x}\rangle.
\end{equation}
For convenience, we define the components in the resonant Bloch frame:
\begin{align*}
    \tilde{s}_{\sigma i}= e^{i\frac{H^0}{\hbar}\tau}s_{\sigma i}e^{-i\frac{H^0}{\hbar}\tau},
\end{align*}
then
\begin{equation}
\label{ch7:sz_ave}
    \langle\psi_f|s_{zi}|\psi_f\rangle=\langle\psi_{0x}|(\tilde{s}_{x i}\cos\varphi -\tilde{s}_{yi}\sin\varphi )|\psi_{0x}\rangle.
\end{equation}

To calculate the correlation, start from:
\begin{equation}
\label{eq:ch7corr_raw}
     \langle\psi_f|s_{zi}s_{zj}|\psi_f\rangle=\langle\psi_{0x}|e^{i\frac{H}{\hbar}\tau}\underbrace{e^{i\frac{\pi}{2}S_y}s_{zi}s_{zj}e^{-i\frac{\pi}{2}S_y}}_{(*)}e^{-i\frac{H}{\hbar}\tau}|\psi_{0x}\rangle.
\end{equation}
Inserting identity operator $\mathbb{I}$:
\begin{align*}
    (*)=e^{i\frac{\pi}{2}S_y}s_{zi}s_{zj}e^{-i\frac{\pi}{2}S_y}=e^{i\frac{\pi}{2}S_y}s_{zi}e^{-i\frac{\pi}{2}S_y}e^{i\frac{\pi}{2}S_y}s_{zj}e^{-i\frac{\pi}{2}S_y}.
\end{align*}
Using Eq. \ref{eq:ch5RotY} again:
\begin{align*}
    (*)=s_{xi}s_{xj}.
\end{align*}
Then Eq. \ref{eq:ch7corr_raw} becomes:
\begin{align*}
     \langle\psi_f|s_{zi}s_{zj}|\psi_f\rangle&=\langle\psi_{0x}|e^{i\frac{H}{\hbar}\tau }s_{xi}s_{xj}e^{-i\frac{H}{\hbar}\tau}|\psi_{0x}\rangle\\
     &=\langle\psi_{0x}|e^{i\frac{H^0}{\hbar}\tau }e^{i\varphi S_z}s_{xi}s_{xj}e^{-i\frac{H^0}{\hbar}\tau}e^{-i\varphi S_z}|\psi_{0x}\rangle.
\end{align*}
Exchange $H^0$ and $S_z$ terms and inserting $\mathbb{I}$ again:
\begin{equation}
     \langle\psi_f|s_{zi}s_{zj}|\psi_f\rangle=\langle\psi_{0x}|e^{i\frac{H^0}{\hbar}\tau }\underbrace{e^{i\varphi S_z}s_{xi}e^{-i\varphi S_z}}_{(*1)}\underbrace{e^{i\varphi S_z}s_{xj}e^{-i\varphi S_z}}_{(*2)}e^{-i\frac{H^0}{\hbar}\tau}|\psi_{0x}\rangle.
\end{equation}
Evaluating $(*1)$ and $(*2)$ with Eq. \ref{eq:Sxofvarphi} again, we obtain:
\begin{equation}
     \langle\psi_f|s_{zi}s_{zj}|\psi_f\rangle=\langle\psi_{0x}|e^{i\frac{H^0}{\hbar}\tau }(s_{xi}\cos\varphi -s_{yi}\sin\varphi )(s_{xj}\cos\varphi -s_{yj}\sin\varphi )e^{-i\frac{H^0}{\hbar}\tau}|\psi_{0x}\rangle.
\end{equation}
Using trigonometry and rearranging terms, the correlation has the form:

\begin{align}
        \label{eq:ch7corrSxy}
   \Rightarrow \mathcal{C}_{ij}^{\perp}\equiv\langle \psi_f| s_{zi}s_{zj}|\psi_f\rangle&=\frac{1}{2}\langle \psi_{0x}| \tilde{s}_{xi}\tilde{s}_{xj}+\tilde{s}_{yi}\tilde{s}_{yj}|\psi_{0x}\rangle \nonumber \\ 
    &+\frac{1}{2}\langle \cos(2\varphi )\rangle\langle \psi_{0x}| \tilde{s}_{xi}\tilde{s}_{xj}-\tilde{s}_{yi}\tilde{s}_{yj}|\psi_{0x}\rangle \\
    &-\frac{1}{2}\langle \sin(2\varphi )\rangle\langle \psi_{0x}| \tilde{s}_{xi}\tilde{s}_{yj}+\tilde{s}_{yi}\tilde{s}_{xj}|\psi_{0x}\rangle.\nonumber
\end{align}

\subsubsection{Zero scattering length limit}
At the magnetic field where the scattering length vanishes, the spin vector evolution is independent of the mean-field interaction: the system evolves under Zeemen precession only. In this section, we derive the prediction for the measured ensemble averaged transverse components of the spin vector and their correlation at $0\,a_0$, including an uncontrolled detuning. In this case, 
\begin{equation}
    \label{eq:ch7H00a0}
    \frac{H^0}{\hbar}\tau=\sum_i\vartheta_{Ei}s_{zi},
\end{equation}
\begin{equation}
\label{eq:vartheta_Ei}
    \text{where }\,\vartheta_{Ei}\equiv\Omega'E_i\tau=-\frac{\delta\omega_x}{\hbar\overline{\omega}_x}\tau \frac{E_i}{E_F}E_F.
\end{equation}
$\vartheta_{Ei}$ is energy dependent, and directly proportional to the evolution time $\tau$.
Calculating the correlation with Eq. \ref{eq:ch7corrSxy} requires all three components, which are evaluated silimarly to Eq.~\ref{eq:Sxofvarphi}: 
\begin{align}
\label{eq:ch70a0comps}
    e^{i\frac{H^0}{\hbar}\tau}s_{xi}s_{xj}e^{-i\frac{H^0}{\hbar}\tau} &= e^{i\frac{H^0}{\hbar}\tau}s_{xi}e^{-i\frac{H^0}{\hbar}\tau}e^{i\frac{H^0}{\hbar}\tau}s_{xj}e^{-i\frac{H^0}{\hbar}\tau}\nonumber\\
    &=[s_{xi}\cos(\vartheta_{Ei})-s_{yi}\sin(\vartheta_{Ei})][s_{xj}\cos(\vartheta_{Ej})-s_{yj}\sin(\vartheta_{Ej})]\nonumber\\
    e^{i\frac{H^0}{\hbar}\tau}s_{yi}s_{yj}e^{-i\frac{H^0}{\hbar}\tau} &= e^{i\frac{H^0}{\hbar}\tau}s_{yi}e^{-i\frac{H^0}{\hbar}\tau}e^{i\frac{H^0}{\hbar}\tau}s_{yj}e^{-i\frac{H^0}{\hbar}\tau}\\
    &=[s_{yi}\cos(\vartheta_{Ei})+s_{xi}\sin(\vartheta_{Ei})][s_{yj}\cos(\vartheta_{Ej})+s_{xj}\sin(\vartheta_{Ej})] \nonumber\\
    e^{i\frac{H^0}{\hbar}\tau}s_{xi}s_{yj}e^{-i\frac{H^0}{\hbar}\tau} &= e^{i\frac{H^0}{\hbar}\tau}s_{xi}e^{-i\frac{H^0}{\hbar}\tau}e^{i\frac{H^0}{\hbar}\tau}s_{yj}e^{-i\frac{H^0}{\hbar}\tau}\nonumber\\
    &=[s_{xi}\cos(\vartheta_{Ei})-s_{yi}\sin(\vartheta_{Ei})][s_{yj}\cos(\vartheta_{Ej})+s_{xj}\sin(\vartheta_{Ej})]\, .\nonumber
\end{align}
Then with some trigonometry:
\begin{align}
    e^{i\frac{H^0}{\hbar}\tau }(s_{xi}s_{xj}+s_{yi}s_{yj})e^{-i\frac{H^0}{\hbar}\tau } = & \,\cos(\vartheta_{Ej}-\vartheta_{Ei})(s_{xi}s_{xj}
    +s_{yi}s_{yj})\nonumber\\
    &+\sin(\vartheta_{Ej}-\vartheta_{Ei})(s_{yi}s_{xj}-s_{xi}s_{yj})\nonumber \\ 
    e^{i\frac{H^0}{\hbar}\tau }(s_{i}s_{xj}-s_{yi}s_{yj})e^{-i\frac{H^0}{\hbar}\tau } = & \,\cos(\vartheta_{Ej}+\vartheta_{Ei})(s_{xi}s_{xj}+s_{yi}s_{yj})\\
    &-\sin(\vartheta_{Ej}+\vartheta_{Ei})(s_{yi}s_{xj}+s_{xi}s_{yj}) \nonumber\\ 
    e^{i\frac{H^0}{\hbar}\tau }(s_{xi}s_{yj}+s_{xj}s_{yi})e^{-i\frac{H^0}{\hbar}\tau } = & \,\cos(\vartheta_{Ej}+\vartheta_{Ei})(s_{xi}s_{yj}+s_{yi}s_{xj})\nonumber\\
    &+\sin(\vartheta_{Ej}+\vartheta_{Ei})(s_{xi}s_{xj}-s_{yi}s_{yj})\, .\nonumber 
\end{align}

Since $s_{xi}|\uparrow_{xi}\rangle=N_i/2|\uparrow_{xi}\rangle$,  $$\langle\psi_{0x}|s_{xi}s_{xj}|\psi_{0x}\rangle=\frac{N_i}{2}\frac{N_j}{2}.$$ When $i=j$, $\langle s_{xi}s_{yj}\rangle<<N_i^2/4$ for $N_i>>1$. For $i\neq j$, $$\langle\psi_{0x}|s_{yi}s_{xj}|\psi_{0x}\rangle=\langle\psi_{0x}|s_{xi}s_{yj}|\psi_{0x}\rangle=0.$$
Using these results, Eq. \ref{eq:ch7corrSxy} can be simplified and the correlation can be written as:
\begin{align}
        \label{eq:ch7corrSxy0a}
     \textit{\textbf{c}}_{ij}^{\perp}\equiv\frac{\langle \psi_f| s_{zi}s_{zj}|\psi_f\rangle}{N_iN_j/4}&=\frac{1}{2}\cos(\vartheta_{Ej}-\vartheta_{Ei}) \nonumber \\ 
    &+\frac{1}{2}\langle \cos(2\varphi )\rangle\cos(\vartheta_{Ej}+\vartheta_{Ei}) \\
    &-\frac{1}{2}\langle \sin(2\varphi )\rangle\sin(\vartheta_{Ej}+\vartheta_{Ei}).\nonumber
\end{align}

\subsection{Data selection}
\label{sec:data}
Because of the unknown distribution of $\varphi $, $\langle \cos(2\varphi )\rangle$ and $\langle \sin(2\varphi )\rangle$ in Eq.~\ref{eq:ch7corrSxy} can vary, leading to unknown fractions of $\tilde{s}_x\tilde{s}_x$, $\tilde{s}_y\tilde{s}_y$, and $\tilde{s}_x\tilde{s}_y$ contributions. However, by intentionally manipulating the distribution of $\varphi $ such that $\langle\cos(2\varphi )\rangle=\langle\sin(2\varphi )\rangle=0$, the measurement result is expected to give the desired $\langle\tilde{s}_x\tilde{s}_x+\tilde{s}_y\tilde{s}_y\rangle$ correlation.

\begin{figure}[hbtp]
\begin{center}
\includegraphics[width=\textwidth]{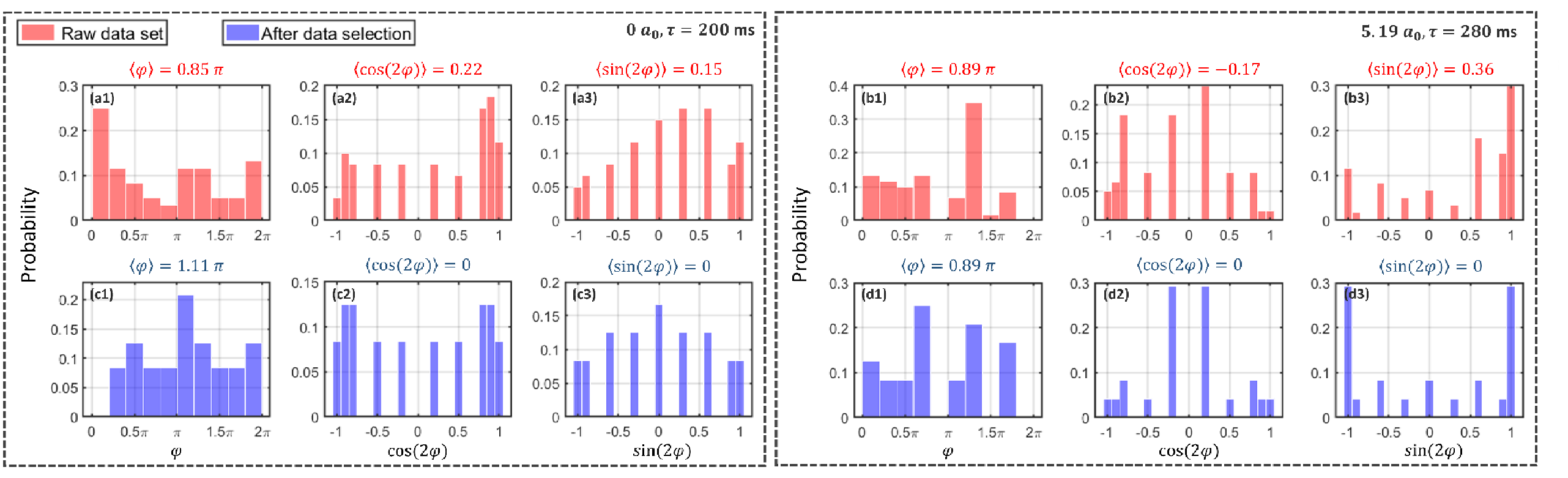}
\end{center}
\caption{Histograms to show examples of distributions of $\varphi$-related quantities before (red) and after (blue) phase selection. (a1,b1,c1) shows the distribution of $\varphi$. (a2,b2,c2) shows the distribution of $\langle \cos(2\varphi)\rangle$. (a3,b3,c3) shows the distribution of $\langle\sin(2\varphi)\rangle$. Left dashed box is for [$0\,a_0,\tau=200$ ms] data set, while right dashed box is for [$5.19\,a_0,\tau=280$ ms].
\label{fig:phiDistr}}
\end{figure}

A data set with the desired $\varphi $ distribution is obtained by data selection. After collecting raw data set with a number of $s^{meas}(x)$ profiles and converting them to $s^{meas}(E)$ profiles using Abel inversion, each $s^{meas}(E)$ is fit with the quasi-classical model to determine fitted $\varphi $ for each shot. Then a collection of pairs of shots are selected from the raw data set, such that in this collection, for each single shot $k$ whose fitted phase is $\varphi^k$, there is another unique single shot $k'$ whose fitted phase $\varphi ^{k'}=\varphi ^k+\frac{\pi}{2}$. $k^{th}$ and $k'^{th}$ shots form a data pair that makes $\langle\cos(2\varphi )\rangle=\langle\sin(2\varphi )\rangle=0$. In this way, for the whole collection of $k$ and $k'$, we obtain $k_{max}$ pairs of single shots and $\langle\cos(2\varphi )\rangle=\langle\sin(2\varphi )\rangle=0$ is guaranteed for the selected $2k_{max}$ shots. Fig.~\ref{fig:phiDistr} shows two examples of histograms of fitted $\varphi$, $\langle \cos(2\varphi)\rangle$ and $\langle\sin(2\varphi)\rangle$ for the raw data set (red, top row) and for the corresponding subset after data selection (blue, bottom row). For the raw data set, the distributions of these $\varphi$-related quantities are pretty random. After the data selection, $\langle \cos(2\varphi)\rangle$ and $\langle\sin(2\varphi)\rangle$ can be constrained to 0.

With this method, for nonzero scattering length, the measured correlation Eq.~\ref{eq:ch7corrSxy} is written as:
\begin{equation}
\label{eq:corr_a_phiAve}
\begin{split}
     \mathcal{C}_{ij}^{\perp}&=\frac{1}{2}\langle \psi_{0x}| \tilde{s}_{xi}\tilde{s}_{xj}+\tilde{s}_{yi}\tilde{s}_{yj}|\psi_{0x}\rangle\\
     \Rightarrow\textit{\textbf{c}}_{ij}^{\perp}&=\frac{4}{N_iN_j}\frac{1}{2}\langle \psi_{0x}| \tilde{s}_{xi}\tilde{s}_{xj}+\tilde{s}_{yi}\tilde{s}_{yj}|\psi_{0x}\rangle.
\end{split}
\end{equation}
For zero scattering length, the measured correlation Eq.~\ref{eq:ch7corrSxy0a} takes the form:
\begin{equation}
    \label{eq:corr_0a0_phiAve}
     \textit{\textbf{c}}_{ij}^{\perp}=\frac{1}{2}\cos(\vartheta_{Ej}-\vartheta_{Ei})
\end{equation}

This data selection method is applied to every data set presented in this work. In the main text, each data point $\mathcal{M}_\perp^2(a,\tau)$ and its error bar in Fig. 2 and 3 are calculated from $k_{max}$ pairs of single shots. First, we make sure $k_{max}$ adopted to calculate error bar is the same for all data sets with [$a,\tau$]. Then to do data selection for each set, we randomly choose 3 pairs of $k$ and $k'$ shots 10 times to obtain 10 $\mathcal{M}_{\perp,k}^2$ values. In the end, we average these 10 $\mathcal{M}_{\perp,k}^2$ values to obtain the ensemble-averaged macroscopic magnetization $\mathcal{M}_\perp^2$. Error bars are calculated by doing statistical standard deviation of these 10 $\mathcal{M}_{\perp,k}^2$ values.

Note that instead of selecting a phase $\varphi$ to be around a specific value (like maximum likelihood estimation), the method presented here enforces a distribution that is flexible, as it employs the mean and deviation of $\varphi$ for the raw data set, resulting in larger fraction of usable data. This method is also tolerant of the fitting model: as long as the fitting model can fit the data qualitatively and includes $\varphi$-dependence correctly, the selected data pairs will have the correct $\varphi_k'=\varphi_k+\frac{\pi}{2}$, ensuring an equal distribution of $\langle s_{xi}s_{xj}\rangle$ and $\langle s_{yi}s_{yj}\rangle$ in $\langle s^{meas}_i s^{meas}_j\rangle$. If the fitting model has a systematic offset $\delta$ in fitting result $\varphi_{k}$ from the real RF detuning $\varphi_{k}^{RF}$, i.e., $\varphi_{k}=\varphi_{k}^{RF}+\delta$, this data selection method, which enforces $\varphi_k'=\varphi_k+\frac{\pi}{2}$, will still give selected data pairs with $\varphi_{k}^{RF'}+\delta= \varphi_{k}^{RF}+\delta+\frac{\pi}{2}$, and therefore $\varphi^{RF'}_k=\varphi_k^{RF}+\frac{\pi}{2}$, the desired distribution.

In contrast, for $\langle \tilde{s}_x\rangle$ or $\langle \tilde{s}_y\rangle$, data selection can only be done by choosing single shots with $\varphi=0$ mod $\pi$ (ensuring $\langle s^{meas}\rangle=\pm\langle\tilde{s}_{x}\rangle$) or $\varphi=\frac{\pi}{2}$ mod $\pi$ (ensuring $\langle s^{meas}\rangle=\pm\langle\tilde{s}_{y}\rangle$). This is inefficient as the selected subset is naturally much smaller than enforcing the pair distribution as described above. Also, the data selection method enforcing a specific $\varphi$ value is highly dependent on the accuracy of fitting model. If the fitting result ends up with a systematic deviation from the real RF phase shift, then the ensemble average of the whole selected data set will have a contribution of the undesired transverse component. Therefore, $\langle \tilde{s}_x\rangle$ or $\langle \tilde{s}_y\rangle$  are not readily obtained by this method.

\subsection{Energy dependent suppression}
\label{sec:ch70a0Corr}
For the experiments presented in this work, the sample is destroyed upon imaging for each shot. Hence all data have slightly varying atom number and cloud size and, therefore, different Fermi energies $E_F$, which determines the maximum Zeeman tuning and the mean-field frequency. The correlation between different energy partitions is presented in units of $E_F$, and higher energy partitions are more sensitive to the variation in $E_F$. In addition, within one shot, there is uncertainty in the measurement of each energy partition $E_i$. This effect arises from the finite energy resolution of the Abel Inversion and spatial resolution of the imaging system. Therefore, after averaging over multiple shots, the magnitude of the measured correlation is suppressed more for higher energy groups. 

To calculate this suppression, we use a normal distribution of $E_F$ and $E_i$:
$$P(E_F|\sigma_{E_F},\mu_{E_F})=\frac{1}{\sigma_{E_F}\sqrt{2\pi}}exp\left[-\frac{1}{2}\left(\frac{E_F-\mu_{E_F}}{\sigma_{E_F}}\right)^2\right];$$
$$P(E_i|\sigma_{E_i},\mu_{E_i})=\frac{1}{\sigma_{E_i}\sqrt{2\pi}}exp\left[-\frac{1}{2}\left(\frac{E_i-\mu_{E_i}}{\sigma_{E_i}}\right)^2\right].$$
Thus, when using Eq.~\ref{eq:corr_0a0_phiAve} for $0\,a_0$ to estimate the suppression coefficient, the measured correlation has the form:
\begin{equation}
    \label{eq:ch7corrSxy0a_avrgphi_2}
     \int_{-\infty}^{\infty}dE_F P(E_F|\sigma_{E_F},\mu_{E_F}) \int_{-\infty}^{\infty}dE_iP(E_i|\sigma_{E_i},\mu_{E_i})\int_{-\infty}^{\infty}dE_jP(E_j|\sigma_{E_j},\mu_{E_j})\frac{1}{2}\cos(\vartheta_{Ej}-\vartheta_{Ei})
\end{equation}
$$\textit{\textbf{c}}_{ij}^\perp\Rightarrow exp\left[-\frac{1}{2}\left(\Omega'\mu_{E_F}\tau\right)^2\left[\left(\frac{\sigma_{E_F}}{\mu_{E_F}}\right)^2\left(\frac{E_i}{E_F}-\frac{E_j}{E_F}\right)^2+\alpha_r^2\left(\frac{E_i}{E_F}+\frac{E_j}{E_F}\right)\right]\right] \textit{\textbf{c}}_{ij}^\perp.$$

This defines an energy-dependent decay factor:
\begin{equation}
\label{eq:Gamma_coeff}
   \Gamma(E_i|\sigma_{E_F}, \alpha_r)\equiv exp\left[-\frac{1}{2}\left(\Omega'\mu_{E_F}\tau\right)^2\left[\left(\frac{\sigma_{E_F}}{\mu_{E_F}}\right)^2\left(\frac{E_i}{E_F}-\frac{E_j}{E_F}\right)^2+\alpha_r^2\left(\frac{E_i}{E_F}+\frac{E_j}{E_F}\right)\right]\right].
\end{equation}

The standard deviation, $\sigma_{E_F}$, and mean, $\mu_{E_F}$, of the Fermi energy $E_F$ are directly extracted from each data set: we calculate $E_F$ for all single shot data from the measured cloud size, $E_F=m\overline{\omega}_x^2\sigma_{Fx}^2/2$, and fit the distribution with a normal probability density function $P(E_F|\sigma_{E_F},\mu_{E_F})$. Typically, for each data set, $\sigma_{E_F}\approx 0.043 \mu_{E_F}$.

$\mu_{E_i}$ is just the mean energy for each energy bin. $\sigma_{E_i}\equiv \alpha_{r}E_F\sqrt{E_i/E_F}$ where $\alpha_{r}$ is found from the energy uncertainty $\delta E_A+\delta E_s$. $\delta E_A$ arises from the number of terms $n_{max}$ adopted to apply Abel inversion as estimated in Eq.~\ref{eq:da}, $\delta E_A\lesssim 0.06\,E_F\sqrt{E_i/E_F}$ for $n_{max}=12$. $\delta E_s$ arises from the finite imaging resolution and is estimated in Eq.~\ref{eq:ds}: $\delta E_s \lesssim 0.03\, E_F\sqrt{E_i/E_F}$. Therefore, the maximum $\alpha_{r}\approx0.09$. In data fitting, the value of $\alpha_r$ is adjusted between 0.06 and 0.09, but for most cases, 0.08 is adopted.

Then the measured correlation after ensemble averaging is predicted to be scaled by $\Gamma$:
\begin{equation}
\label{eq:Gamma_coeff_timesC}
 \Gamma(E_i|\sigma_{E_F}, \alpha_r)\,\textit{\textbf{c}}_{ij}^\perp,
\end{equation}
with \textit{\textbf{c}}$_{ij}^\perp$ being the correct correlation between group $i$ and $j$.

\subsubsection{Testing $\Gamma(E_i|\sigma_{E_F},\,\alpha_r)$}
With the parameters $\sigma_{E_F}$ and $\alpha_r$ estimated, $\Gamma(E_i|\sigma_{E_F},\,\alpha_r)$ can be calculated and examined quantitatively by comparing the data to quasi-classical model prediction with $\Gamma$ included. Since the lowest energy group always has the largest atom number, providing the largest signal, we calculate the correlation between particles in the first (lowest) energy group and in all other energy groups, \textbf{\textit{c}}$_{ij}^\perp$, to test the validity of suppression coefficient $\Gamma(E_i|\sigma_{E_F},\,\alpha_r)$.

\begin{figure}[hbtp]
\begin{center}
\includegraphics[width=0.85\textwidth]{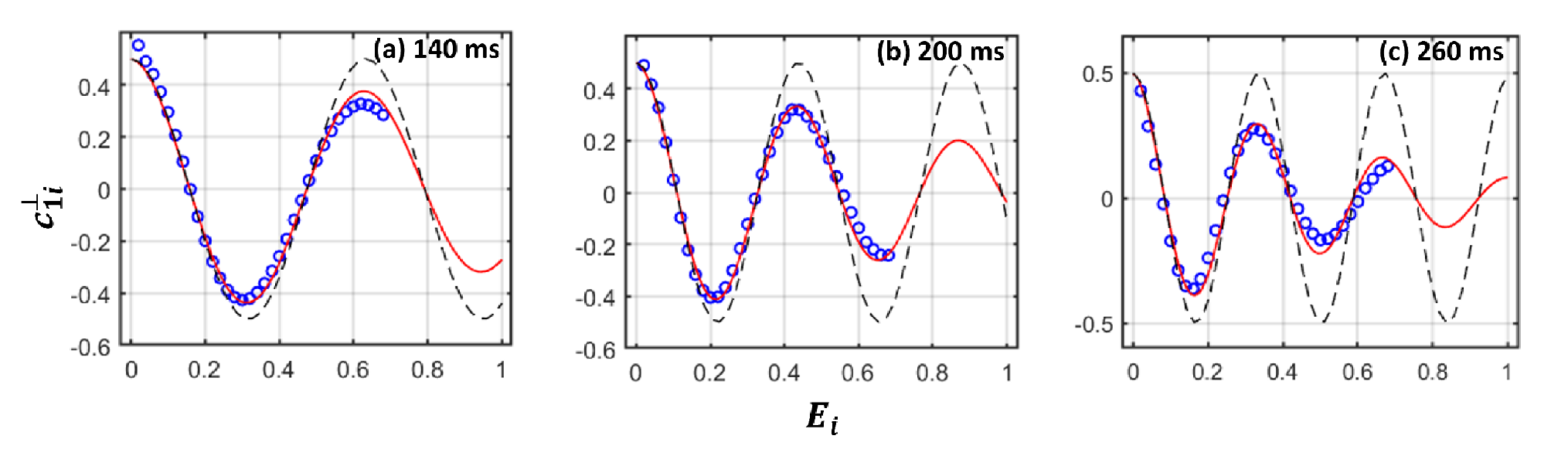}
\end{center}
\caption{Comparing model with suppression factor $\Gamma$ (red curve) and without (black dashed curve) and experimental data (blue circles) for the normalized correlation \textit{\textbf{c}}$_{1j}^\perp$ at different evolution times at the zero-crossing magnetic field. Blue circles are obtained by averaging over 30 single shot data. Black dashed curve is the exact solution given in Eq. \ref{eq:corr_0a0_phiAve}. Red curve is the adjusted model, which includes the shot-to-shot variation in $E_F$ and the finite energy resolution. In this figure, only the lowest $70\%$ energy bins are adopted in data analysis.
\label{fig:ch72halfPi0a0corr}}
\end{figure}

The first test is implemented at zero scattering length. In this case, there is no mean-field interaction. Therefore the model prediction is very reliable as there is no approximation in the Zeeman precession term. Fig. \ref{fig:ch72halfPi0a0corr} shows the result for zero scattering length case. Blue circles are the correlations obtained by averaging over about 30 single-shot data. Black dashed curves are the exact solutions predicted by Eq. \ref{eq:corr_0a0_phiAve}. Note that this curve has a sinusoidal shape and the oscillation amplitude stays the same as it goes from low to high energy groups, while the data only has the same amplitude as the model at lower energy groups and becomes smaller and smaller compared to the model for higher energy groups. Red curves show the $\Gamma$ scaled model, Eq.~\ref{eq:Gamma_coeff_timesC}. As shown in this figure, with the variation in $E_F$ and energy resolution included, the model and data are in quantitative agreement. The oscillation frequency of the data, which is determined by $\vartheta_{Ei}$ defined in Eq. \ref{eq:vartheta_Ei}, is in agreement with both predictions. This confirms that the numerical value of $\Omega'$ adopted in our model is correct.

\begin{figure}[hbtp]
\begin{center}
\includegraphics[width=0.65\textwidth]{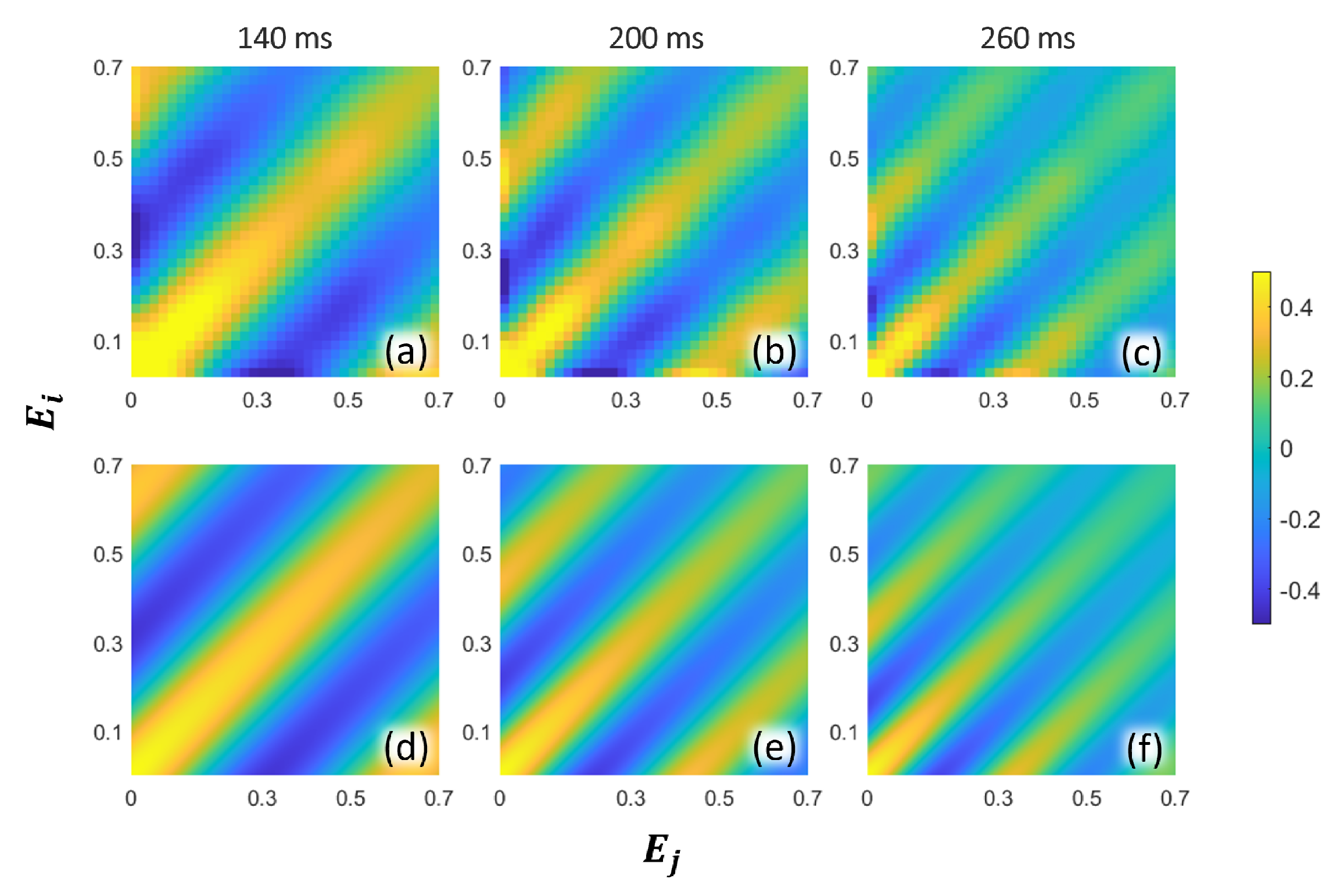}
\end{center}
\caption{Normalized correlation \textit{\textbf{c}}$_{ij}^\perp$ at different evolution times at $0\,a_0$. Top row is calculated from data. Bottom row is predicted by the model using Eq. \ref{eq:corr_0a0_phiAve} and Eq. \ref{eq:Gamma_coeff_timesC}. All panels share the same color bar. In this figure, only the lowest $70\%$ of the energy bins are adopted in data analysis.
\label{fig:0a0_sisj_supp}}
\end{figure}

Fig. \ref{fig:0a0_sisj_supp} visualizes the correlation between all energy partitions \textbf{\textit{c}}$_{ij}^\perp$ to compare model and data. The top row shows data and bottom row shows the model calculated with Eq. \ref{eq:Gamma_coeff_timesC}. This figure confirms the agreement between the adjusted model and data.

The agreement between the model with $\Gamma(E_i|\sigma_{E_F},\alpha_r)$ included and data for the experiment conducted at the zero-crossing suggests that the calculated energy-dependent suppression coefficient $\Gamma$ predicts the decay in correlation because of the multi-shot average and finite energy resolution. To consolidate this idea and apply it to all data analysis, we do the same comparison with the data obtained at nonzero scattering length, $5.19\, a_0$. In such a situation, the exact solution Eq. \ref{eq:corr_0a0_phiAve} is not valid anymore. To predict the correlation, modeling $\tilde{s}_{xi}$ and $\tilde{s}_{yi}$ is required. We extract the experimental parameters $\sigma_{Fx}$ and $N$ from each shot to estimate $\tilde{s}_{xi}$ and $\tilde{s}_{yi}$ with the quasi-classical spin model, then calculate $\tilde{s}_{xi}\tilde{s}_{xj}$ and $\tilde{s}_{yi}\tilde{s}_{yj}$. Thus $\langle \psi_{0x}| \tilde{s}_{xi}\tilde{s}_{xj}|\psi_{0x}\rangle$ and $\langle \psi_{0x}| \tilde{s}_{yi}\tilde{s}_{yj}|\psi_{0x}\rangle$ are obtained by averaging over multiple predictions and \textit{\textbf{c}}$_{ij}^\perp$ is predicted using Eq. \ref{eq:ch7corrSxy}. 

The model in  Eq. \ref{eq:ch7corrSxy} is not written in terms of $E_F$ or $E_i$ explicitly, so it is harder to include shot-to-shot variation in $E_F$ and finite resolution of $E_i$ in the model rigorously. However, a rough estimation can be made by applying the same exponential decay factor in Eq. \ref{eq:Gamma_coeff} to Eq. \ref{eq:corr_a_phiAve}. Thus, after the data selection, the measured correlation is predicted with:
\begin{equation}
\label{eq:ch7corrSxy_2}
    \Gamma(E_i|\sigma_{E_F}, \alpha_r) \frac{1}{2}\langle \psi_{0x}| \tilde{s}_{xi}\tilde{s}_{xj}+\tilde{s}_{yi}\tilde{s}_{yj}|\psi_{0x}\rangle. 
\end{equation}
$\mu_{E_F}$ and $\sigma_{E_F}$ are obtained by fitting normal probability density function to $E_F$ distribution for each data set. $\alpha_r$ is the same value as calculated for $0\,a_0$ case: $\alpha_r=0.09$. 

\begin{figure}[hbtp]
\begin{center}
\includegraphics[width=0.85\textwidth]{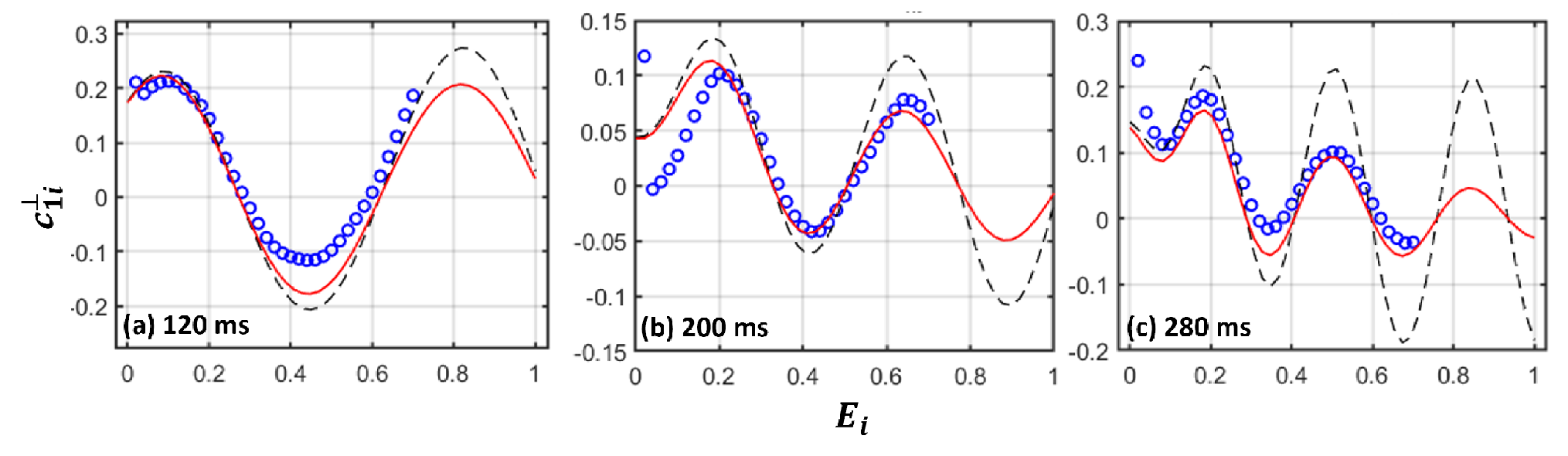}
\end{center}
\caption{Comparing model with suppression factor $\Gamma$ (red curve) and without (black dashed curve) and experimental data (blue circles) for the normalized correlation \textit{\textbf{c}}$_{1j}^\perp$ at different evolution times at $5.19\,a_0$. Blue circles are obtained by averaging over 30 single shot data. Black dashed curve is the exact solution given in Eq. \ref{eq:corr_a_phiAve}. Red curve is the adjusted model, which includes the shot-to-shot variation in $E_F$ and the finite energy resolution. In this figure, only the lowest $70\%$ energy bins are adopted in data analysis.
\label{fig:ch75a0halfPisz1szi}}
\end{figure}

The result of correlation between first and all other energy groups predicted by this adjusted model is shown in red curves in Fig. \ref{fig:ch75a0halfPisz1szi}. In this figure, blue circles are data, black dashed curves are the raw model without $E_F$ variation and finite energy resolution. The agreement between data and the raw model is qualitative: the oscillation frequencies of correlation for the data and model are the same, but the amplitude of correlation calculated from data is obviously smaller than the raw model prediction. In contrast, the model including $E_F$ variation and $E_i$ uncertainty fits the data much better.

By comparing the ensemble-averaged \textit{\textbf{c}}$_{ij}^\perp$ over the data set after data selection to the quasi-classical model with an energy-dependent coefficient $\Gamma(E_i|\sigma_{E_F},\alpha_r)$ included, we confirm that the numerical implementation of $\Gamma(E_i|\sigma_{E_F},\alpha_r)$ is consistent with data as described in this section. Therefore, for all \textit{\textbf{c}}$_{ij}^\perp$ plots presented in this work, the calculated \textit{\textbf{c}}$_{ij}^\perp$ is shown after being multiplied by $\Gamma^{-1}(E_i|\sigma_{E_F},\alpha_r)$ to restore the suppressed signal to the correct multi-shot average. The value of $\alpha_r$ adopted varies from 0.06 to 0.09 for fitting purposes. Note that, in $\mathcal{M}_\perp^2$ calculation, $\Gamma^{-1}(E_i|\sigma_{E_F},\alpha_r)$ is not needed since the double summation $\sum_{i,j}\mathcal{C}_{ij}^\perp$ is implemented for every single shot, avoiding the suppression because of average. In the end, the $\mathcal{M}_\perp^2$ for the selected data set is obtained by averaging over that for all single shots.

\subsection{Quantifying uniformity of transverse spin correlations in energy space}

By observing the spread/localization behavior of microscopic transverse spin correlation in energy space, our work opens new ways to study the phase transition in a many-body system. One method to quantify the structure of the surface plots is to measure how uniformly the transverse correlation has spread across all energy group pairs for the system at evolution time $\tau$ with interaction strength $\zeta$. This section describes the details of the calculation method for this uniformity.

\textit{\textbf{c}}$_{ij}^\perp$ of energy pairs $E_i,E_j\in [0, 0.7]E_F$ are adopted, as the larger energy group contains very few atoms, the signal-to-noise ratio is low. Therefore in the data analysis, the lowest 35 energy groups out of 50 were adopted. Then the \textit{\textbf{c}}$_{ij}^\perp$ being analyzed is a $35\times 35$ matrix with $i,j=1,2,...,35$. Each $i^{th}$ row of this matrix as a function of $j$ will display an oscillating curve similar to those Fig. \ref{fig:ch72halfPi0a0corr} and \ref{fig:ch75a0halfPisz1szi}. Then, a MATLAB Savitzky-Golay filter is applied to data in each row to reduce noise, resulting in a smoothed curve described by a function $f_{i,sm}^\perp(j)$ for $i^{th}$ row with $j$ being variable. Then we find the row number $i=m$, which gives the maximum value of $f_{i,sm}^\perp(j)$ across all $j$ for all $i$
. This way, the center of the highest correlated region for this surface plot is located at the $m^{th}$ energy group. 

Using 35 values of \textit{\textbf{c}}$_{mj}^\perp$, we find the numerical gradient at each interior data point ($j\in[2, 34]$):
\begin{equation}
    \nabla\mathbb{\textit{c}}_{mj}^\perp \equiv \frac{1}{2}\times (\mathbb{\textit{c}}_{m,j+1}^\perp-\mathbb{\textit{c}}_{m,j-1}^\perp).
    \label{eq:grad_1}
\end{equation}
For the data points at two edges of the array ($j = 1,35$), the gradient is calculated by:
\begin{equation}
    \nabla\mathbb{\textit{c}}_{m1}^\perp \equiv \mathbb{\textit{c}}_{m2}^\perp-\mathbb{\textit{c}}_{m1}^\perp,
        \label{eq:grad_2}
\end{equation}
\begin{equation*}
    \nabla\mathbb{\textit{c}}_{m,35}^\perp \equiv \mathbb{\textit{c}}_{m,35}^\perp-\mathbb{\textit{c}}_{m,34}^\perp.
\end{equation*}

With Eq. \ref{eq:grad_1} and \ref{eq:grad_2}, the mean of absolute values of \textit{\textbf{c}}$_{mj}^\perp$ gradient is calculated and defined as $\mathcal{D}_m$: 
\begin{equation}
    \mathcal{D}_m\equiv\frac{1}{35}\sum_{j=1}^{35}|\nabla\mathbb{\textit{c}}_{mj}^\perp|,
\end{equation}

\begin{figure}[hbtp]
\begin{center}
\includegraphics[width=0.7\textwidth]{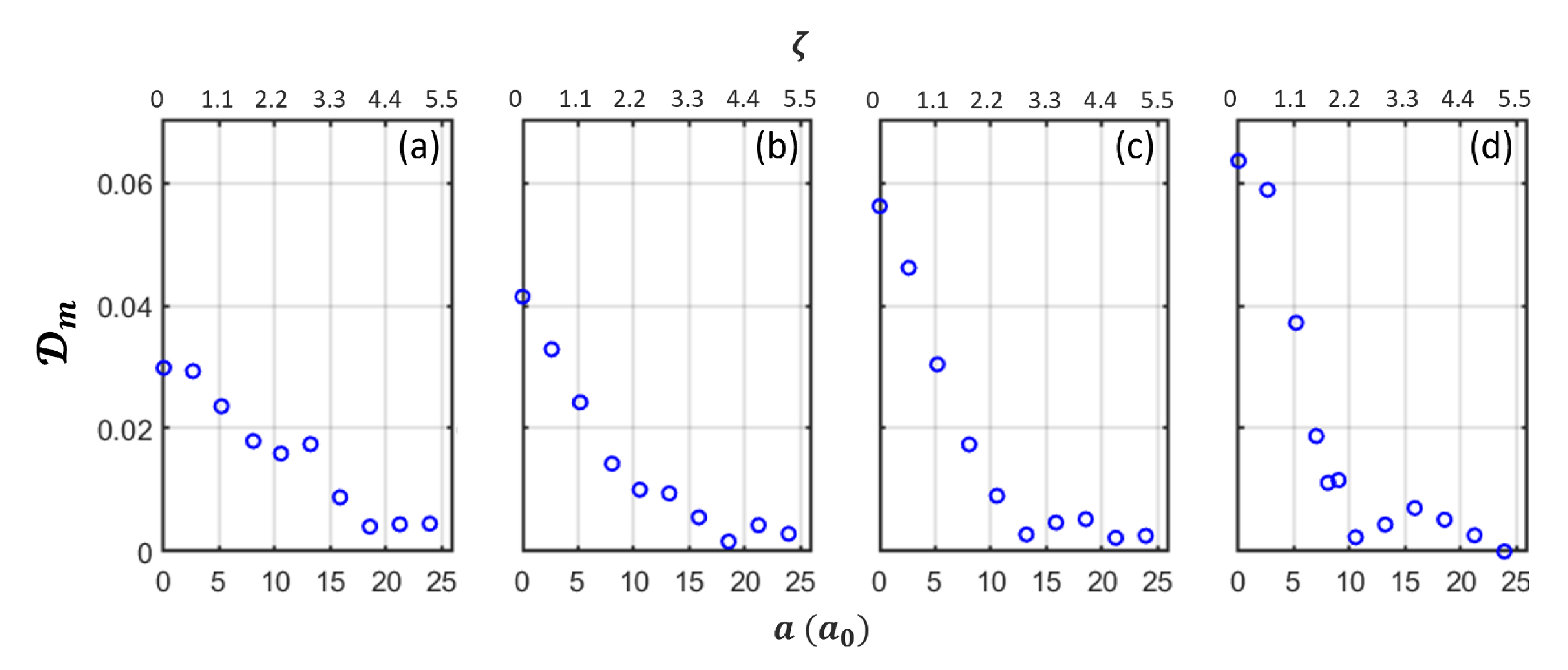}
\end{center}
\caption{$\mathcal{D}_m$ as a function of interaction strength $\zeta$ at four evolution times (a) $\tau=80$ ms, (b) $\tau =120$ ms, (c) $\tau = 160$ ms, and (d) $\tau = 200$ ms.
\label{fig:gradMax_vary_a}}
\end{figure}

$\mathcal{D}_m$ calculated from the energy resolved transverse spin correlation measurement also serves the purpose of detecting the phase transition of the system. Similar to the surge observed in $\frac{1}{2}\mathcal{M}_\perp ^2$ measurement as interaction strength increases(Fig. 3 in main text), $\mathcal{D}_m$ value drops abruptly where the phase transition occurs (Fig. \ref{fig:gradMax_vary_a}).

\subsection{Spin-locking in the spin vector picture}

In this work, it is observed that the system undergoes a transition to a ferromagnetic phase as interaction strength increases, shown by a sharp rise in $\frac{1}{2}\mathcal{M}_\perp^2$ (Fig. 3 in the main text). The spin vector picture provides a physical illustration of this transition. Recall that, by definition, $\mathcal{M}^2_\perp=S_x^2+S_y^2$. Thus, the magnetization is related to the dispersion of the spin vector in the $xy$-plane: the more spins cluster, the larger magnitude $\mathcal{M}^2_\perp$ has. This can be considered as a spin-locking effect. 

\begin{figure}[hbtp]
\begin{center}
\hspace*{-0.15in}
\includegraphics[width=0.65\textwidth]{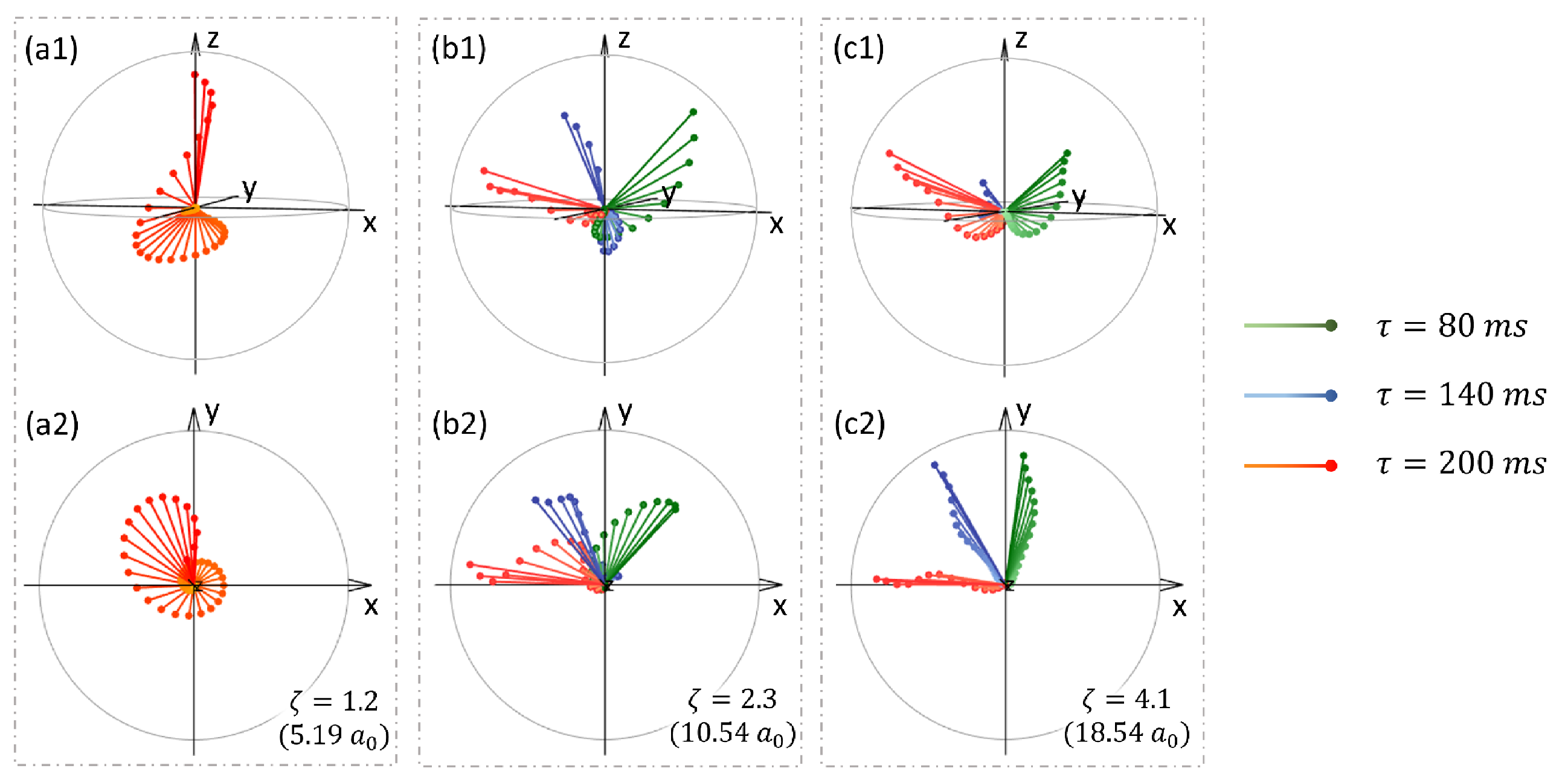}
\end{center}
\caption{Modeled spin vectors in the Bloch resonant frame for different energy partitions (longer segments represent spin vectors with lower energy and vice versa). (a1,a2) are spin vectors with $\zeta=1.2$ at $200$ ms. (b1,b2,c1,c2) are spin vectors at different $\tau$ with $\zeta=2.3$ and $4.1$ respectively. (a2,b2,c2) are the top views of (a1,b1,c1). Red, blue, and green segments are spins at $200$, $140$, and $80$ ms respectively. \label{fig:sphere1}}
\end{figure}

Fig.~\ref{fig:sphere1} depicts the spin-locking phenomenon using the quasi-classical spin model, where $x$, $y$, and $z$ denote axes in the Bloch frame. (a1,a2) shows the spin vectors with different energies after evolving for $200$ ms with a small interaction strength $\zeta=1.2$. (a2) is the top view of (a1) and clearly shows spin vectors in different energy partitions are largely spread out over all four quadrants in the $xy$-plane. In the microscopic correlation picture, spins with the same or opposite azimuthal angles are correlated, and positive and negative single-pair correlations tend to cancel each other, leaving a weak magnetization after double summation over all energy partitions, corresponding to low $\mathcal{M}^2_\perp$ value in Fig. 3(d) in main text. In contrast, (c1,c2) demonstrate a spin-locked state, with $\zeta=4.1$ after evolving for 80 ms (green), 140 ms (blue), and 200 ms (red). For all three evolution times, the spin vectors in all energy partitions tend to congregate. In this situation, spins in all energy partitions are strongly and positively correlated, resulting in a highly magnetized state, in agreement with Fig. 3 in main text  for $\zeta=4.1$. (b1,b2) shows an intermediate stage between (c1,c2) and (a1,a2): the spin vectors have not formed a bundle at $\tau=80$ ms (green), but start showing this trend at $\tau=140$ ms, (blue) and $200$ ms (red). Further, as interaction strength increases, $s_{zi}$ also tends to cluster, with $\langle S_z^2\rangle$ becoming small as $\langle S_x^2+S_y^2\rangle$ increases. 

\end{document}